\documentclass{amsart}
\usepackage{float}
\usepackage{multirow}
\usepackage{enumitem}
\usepackage{amsmath,amsfonts,amssymb}
\usepackage{bm}
\usepackage{hyperref}
\usepackage{xcolor}
\usepackage{geometry}
\usepackage[ruled]{algorithm2e}
\usepackage[noend]{algpseudocode}

\SetCommentSty{mycommfont}

\usepackage{subfig}
\theoremstyle{definition}

\theoremstyle{remark}

\numberwithin{equation}{section}
\newcommand*{\set}[1]{\left\{#1\right\}}

\newcommand*{\norm}[1]{\left\Vert#1\right\Vert}
\newcommand*{\opnorm}[1]{{\left\vert\kern-0.25ex\left\vert\kern-0.25ex\left\vert #1 
		\right\vert\kern-0.25ex\right\vert\kern-0.25ex\right\vert}}

\newcommand*{\probref}[1]{\hyperref[#1]{Problem \ref{#1}}}
\newcommand*{\lemref}[1]{\hyperref[#1]{Lemma \ref{#1}}}
\newcommand*{\thmref}[1]{\hyperref[#1]{Theorem \ref{#1}}}
\newcommand*{\exref}[1]{\hyperref[#1]{Example \ref{#1}}}
\newcommand*{\defnref}[1]{\hyperref[#1]{Definition \ref{#1}}}
\newcommand*{\algoref}[1]{\hyperref[#1]{Algorithm \ref{#1}}}
\newcommand*{\figref}[1]{\hyperref[#1]{Figure \ref{#1}}}
\newcommand*{\tableref}[1]{\hyperref[#1]{Table \ref{#1}}}

\newcommand*{\secref}[1]{\hyperref[#1]{Section \ref{#1}}}

\DeclareMathOperator*{\argmin}{argmin}

\newcommand{\mycomment}[1]{}
\usepackage{todonotes}
\newcounter{todocounter}

\begin{document}
	
\title{Boosting Convolutional Neural Networks' Protein Binding Site Prediction Capacity using SE(3)-invariant transformers, transfer learning and homology-based augmentation}
\date{\today}
\author{Daeseok Lee}\email{daelee@deargen.me}
\author{Jeunghyun Byun}\email{jhbyun@deargen.me}
\author{Bonggun Shin}\email{bonggun.shin@deargen.me}
\begin{abstract}
Figuring out small molecule binding sites in target proteins, in the resolution of either pocket or residue, is critical in many virtual and real drug-discovery scenarios. Since it is not always easy to find such binding sites based on domain knowledge or traditional methods, different deep learning methods that predict binding sites out of protein structures have been developed in recent years. Here we present a new such deep learning algorithm,  that significantly outperformed all state-of-the-art baselines in terms of the both resolutions---pocket and residue. This good performance was also demonstrated in a case study involving the protein human serum albumin and its binding sites. Our algorithm included new ideas both in the model architecture and in the training method. For the model architecture, it incorporated SE(3)-invariant geometric self-attention layers that operate on top of residue-level CNN outputs. This residue-level processing of the model allowed a transfer learning between the two resolutions, which turned out to significantly improve the binding pocket prediction. Moreover, we developed novel augmentation method based on protein homology, which prevented our model from over-fitting. Overall, we believe that our contribution to the literature is twofold. First, we provided a new computational method for binding site prediction that is relevant to real-world applications, as shown by the good performance on different benchmarks and case study.  Second, the novel ideas in our method---the model architecture, transfer learning and the homology augmentation---would serve as useful components in future works. 
\end{abstract}
	
\maketitle

\section{Introduction}

In structure-based drug discovery, the knowledge of ligand binding sites (hereafter binding sites) on target proteins is crucial. It can aid \textit{rational drug design} (\cite{SiteMap}, \cite{orphan_gpcr}, \cite{galectin_10}) and is required for \textit{in-silico} methods such as docking (\cite{large_scale_docking}, \cite{orphan_gpcr}). Such knowledge of binding sites can be attained by analyses of experimental structures of the target protein in complex with ligands. However, if no such structure is at one's disposal, it may be necessary to rely on computational means in order to identify the binding sites. 

In general, this computational task of \textbf{Binding Site Prediction (BSP)} can be regarded as composition of two sub-tasks: (1) \textbf{Binding Site Detection (BSD)} and (2) \textbf{Binding Residue Identification (BRI)}. 

Firstly, BSD aims to identify the binding sites in a coarse-grained manner and score their druggability. A successful detection of highly druggable binding sites can aid medical chemists in many ways when designing  better drug compounds. For example, the medical chemists can draw valuable insights in improving drug compounds' binding affinity or physical properties by examining the receptor structure at the potential binding site (\cite{SiteMap}). Also, preparing a suitable binding site is the first step in any virtual structure-based drug discovery pipeline (\cite{large_scale_docking}). 

Secondly, BRI aims to identify residues in a given binding site that plays key roles in interactions with ligands. Identification of the key residues has been pursued in many previous research papers, due to its importance in rational drug design (\cite{orphan_gpcr}, \cite{galectin_10}, \cite{kinase_database}).  In particular, it has several applications in virtual structure-based drug discovery. 
For example, \textit{structural pharmacophore} features can be selected based on the identified key residues (\cite{orphan_gpcr}, \cite{hs-pharm}), and docking results can be prioritized according to whether the docked molecule has favored interactions with the key residues (\cite{large_scale_docking}, \cite{orphan_gpcr}).

In this paper, we will focus in particular on the structure-based Deep Learning methods to tackle the BSP problem. This choice reflects two recent trends. Firstly, Deep Learning has been widely adopted for BSP (\cite{exploring_bsp}), and has shown good performance. Secondly, it has become easier to identify protein structures as a result of (1) the rapidly accumulating experimental data in databases such as PDB, and (2) advances in Deep Learning methods e.g. Alphafold (\cite{alphafold}).

Recent structure-based Deep Learning methods for BSP are predominantly based solely on 3D CNN architectures that operate on grid-shaped inputs. This results in several limitations, in terms of the performance of two sub-tasks explained previously.

Firstly, the way the CNN-based methods aggregate the local (or short-range) information to recognize the global patterns may be sub-optimal. They achieve this through either clustering algorithms applied on top of the CNN outputs, or through the CNN layers themselves. In the first case, the clustering algorithm could be replaced by highly parameterized modules such as Neural Networks. These can potentially outperform the clustering algorithm, since many parameters can be optimized for the given task. In the second case, the deep layers of convolution operations may suffer from the problem of ``long-term dependency". This means, since a convolution layer's operation is only local, a deep hierarchy of convolutions must be applied in order to allow a neuron to have a receptive field large enough to capture the global patterns. This long-term dependency is known to impede the training (\cite{grad_flow_rnn}). 
This problem can be resolved if we adopt a neural network architecture whose operation is not local in nature.

Secondly, these grid-based models do not directly operate on the protein residues. Therefore, they need an ad-hoc conversion of outputs to obtain predictions about the binding residues. For example, they posit that an atom close to a predicted 3D point (e.g. a predicted voxel center) is part of a binding residue (\cite{DeepSite}, \cite{Kalasanty}).  This may lead to sub-optimal BRI performances, since the loss function used to train the model do not compare the ground truth with the final output, but with the intermediate output before the conversion. For a better BRI performance, having a parameterized model that directly outputs the residue-level predictions would be preferable.

% For example, even if a residue is far from any ligands, it may be close to a voxel center that is close to a ligand, which may induce the model to erroneously predict the residue to be a binding site residue. 

To resolve these problems, we devised a model that has geometric attention layers that operate on top of residue-level CNN outputs. Our model is composed of two modules dedicated to the sub-tasks BSD and BRI respectively. Both modules (1) divide local surroundings around each protein residue into grids, (2) process the grid features in parallel using a CNN model to obtain residue-level local features, (3) use the geometric self-attention layers to update the features, and (4) compute the final reductions to produce subtask-specific outputs. 

To further improve our model's performance, we devised several additional elements---transfer learning, SE(3)-invariance, and augmentation. All these elements are intimately connected to our model's architecture. 
%Therefore, the fact that our model can accompany these elements is an additional benefit.

Firstly, we applied the transfer learning between the two modules. More specifically, the BSD module was not trained from scratch but a portion of its parameters was initialized from the trained parameters of the BRI module. This was meant to facilitate the learning process of the BSD module, which may suffer from a relative lack of data.

Secondly, in order to improve the robustness of our model's predictions, we adapted the model to be SE(3)-invariant. In other words, our model is invariant to rotation and translation of the input structure (\cite{se3_transformers}). Besides using an SE(3)-invariant attention mechanism (\cite{alphafold}) at the outset, we made the grid featurization process SE(3)-invariant as well. This was achieved by aligning the axes of the grids with a specific orientation. This is similar to the grid alignment method used in \cite{DeepSurf}. 
While the grids aligned with that method is not completely deterministic (as one degree of freedom remains), the grids aligned in our method do and hence achieve the full SE(3)-invariance.

Lastly, we devised the data augmentation techniques that can be used along with our model. This was necessary because existing augmentation methods based on translations or rotations have no effect on training the SE(3)-invariant model. In particular, we came up with a novel augmentation method based on protein homology search and sequence alignment. Protein homology has been utilized for the BSP problem (\cite{findsite}, \cite{3dligandsite}, \cite{cofactor}, \cite{tmsite_ssite}), but to the best of our knowledge, we were the first to use it as an augmentation method for the same problem.

The resulting method achieved significant performance gains over previous methods when evaluated on various BSP datasets, both in terms of BSD and BRI. A BSD metric increased by 3.8\% on average, and a BRI metric increased by 16.9\% on average. 

Through an ablation study, we showed that all the components of our model---the model architecture, transfer learning, SE(3)-invariance, and augmentation---made significant contributions, with a few exceptions for the BSD metric. At the end of the experiment section, we provided potential reasons for such exceptions.

Finally, we performed a case study on \textit{human serum albumin} to show the effectiveness of our model in real world applications. We based our case study on two previous studies on the binding sites of human serum albumin (\cite{hsa_oncology_drugs}, \cite{hsa_survey}), and examined how well our method could make predictions that are compatible with the studies. The promising results showed the potential usefulness of our method in in-silico drug discovery. 
%In particular, since we let our model make predictions based on a structure computed by Alphafold (\cite{alphafold}, \cite{alphafold_human_proteome}) during the case study, our model is potentially useful on proteins for which no experimental structures are known. 

To summarize, our contributions in this paper are: 
\begin{itemize}
\item We developed a new SE(3)-invariant deep learning model for BRI and BSD that combine CNN with the geometric self-attention layers.  
\item We developed data augmentation methods, in particular homology augmentation, that can be used when training our model. 
\item We found that our new model, trained with the proposed data augmentation methods, achieved significant performance gains over state-of-the-art deep learning methods for BRI and BSD. 
\item By an ablation study, we found that all elements of our method contributed significantly to BRI performance. 
\item By an ablation study, we examined which elements contributed unambiguously to BSD performance. For those that did not, we provided possible explanations. 
\end{itemize}

\section{Related Works}\label{section:related_works}
In this section, we discuss (1) existing BSP methods, focusing on the traditional and the deep learning methods, and (2) a similar problem of predicting ligand-specific binding sites. 
\subsection{Traditional BSP methods}
\subsubsection{Probe-based Methods}
These methods use a fixed set of small molecules called "probes" to determine the binding sites in a query protein (\cite{q-sitefinder}, \cite{sitehound}, \cite{ftsite}). Specifically, they place the probes at different positions on the surface of the protein, and calculate the physical energy at the positions. The low-energy positions are predicted to be the  potential binding sites. 

\subsubsection{Geometry-based Methods}
These methods rely on 3D geometric characterization of binding sites to detect them. (\cite{castp}, \cite{Fpocket}, \cite{mspocket})

One example is Fpocket (\cite{Fpocket}), that tries to find concave regions of appropriate sizes on the protein surface. It does so by approximating the local curvatures by radii of alpha spheres, which are spheres with four heavy atoms on them but no heavy atom inside. More specifically, it finds all alpha spheres within a radius range, clusters them, and filters them according to the number of constituent alpha spheres to produce a binding pocket. 

Although Fpocket typically produces an excessive number of binding pockets, it has relatively good recall (96.4\% on scPDB v.2017, according to \cite{scPDB}). Therefore, there are ML-based algorithms (\cite{p2rank}, \cite{Deeppocket}) that make use of it as a means to generate initial candidate binding sites. Our method employs the same strategy.  

%Fpocket is a widely used BSP software, the algorithm of which is solely based on the geometry of the heavy (non-hydrogen) atoms in a protein. It is based on the assumption that binding sites tend to be \textit{cavities}, hollow spaces surrounded by atoms, of appropriate sizes. Although it typically produces many binding site candidates including those that are not actual binding sites, it is known that it has very good recall (96.4\% on scPDB v.2017, according to \cite{scPDB}).
\subsubsection{Template-based Methods}\label{subsubsection:template-based_methods}

These methods predict binding sites of a query protein based on the \textit{templates}, which are other similar proteins whose binding sites are already known (\cite{findsite}, \cite{3dligandsite}, \cite{cofactor}, \cite{tmsite_ssite}). A portion of the query protein is regarded as a binding site if it resembles binding sites of templates either sequentially or structurally. 

For example, the authors of \cite{tmsite_ssite} suggested combining two template-based approaches, one based on substructure comparison (TM-SITE) and the other on sequence profile alignment (S-SITE). 
TM-SITE works as follows: 
\begin{enumerate}
\item Putative binding pockets are identified in the query protein by relying on an external software (\cite{concavity}).
\item For each putative binding pocket, the template binding sites similar to it are collected as putative templates.
The similarity measure is based on both structural and sequential comparisons.
\item The ligands in the putative templates are projected to the binding pocket. 
\item A consensus voting by the projected ligands determines whether the residues in the binding pocket are in the binding site or not.
\end{enumerate}
On the other hand, S-SITE works as follows:
\begin{enumerate}
    \item The query protein sequence is aligned with the template sequences based on their position-specific scoring matrices (PSSM) profiles and secondary structure information. 
    \item Templates with the highest alignment \textit{quality scores} are chosen as the putative templates. 
    \item
    A consensus voting by the templates determines whether the residues in the query sequence are in the binding site. 
\end{enumerate}

Our homology-based augmentation algorithm is inspired by these methods. The overall flow of it resembles that of TM-SITE, and the use of global sequential alignment is shared by S-SITE. However, our aim in applying the algorithm is not to make a final prediction of the model, but rather to augment the training dataset.

\subsection{deep learning based BSP Methods}\label{subsection:dl_based-related_works}

\subsubsection{DeepSite}
DeepSite(\cite{DeepSite}) predicts the binding sites by using a 3D CNN model and a clustering algorithm. The inference steps of DeepSite are as follows: (1) it generates points spanning the entire protein-occupied 3D space, (2) predicts the \textit{ligandability} of the points using the CNN model computed on a 3D grid centered at the points, and (3) clusters the ligandable points to produce binding sites.

\subsubsection{DeepSurf}
DeepSurf(\cite{DeepSurf}) has the same overall procedure with DeepSite(\cite{DeepSite}), but it uses more sophisticated approaches in several aspects. More specifically, it tries to improve (1) the generation of initial points, (2) the formation of input grids, and (3) the architecture of the 3D CNN model. The initial points are sampled on the Solvent Accessible Surface (SAS) of the protein, rather than the entire span of the protein. Then, the axes of the grids formed at those points are not arbitrarily oriented, but one axis is set to be the normal vector of the SAS. Finally, rather than using a plain CNN model, they used 3D equivalents of ResNet and Bottleneck ResNet (\cite{resnet}).

\subsubsection{Kalasanty}

Kalasanty (\cite{Kalasanty}) tries to solve BSP by viewing it as a 3D \textit{image segmentation} problem. Therefore, it uses a 3D equivalent of the U-net model (\cite{U-net}), which was originally developed for 2D images. It applies the U-net model to large grids that cover most of the query proteins, and outputs the connected components consisting of positively predicted voxels as binding sites. 

\subsubsection{Deeppocket}
Similar to our proposed method, Deeppocket (\cite{Deeppocket}) relies on the binding site candidates generated by Fpocket. It has separate detection and segmentation models, where the former is a plain 3D CNN model, and the latter is a U-net model similar to the one used in Kalasanty. The detection model is used to rank the binding site candidates generated by Fpocket, and the segmentation model is used to segment the 3D voxels centered at the top-ranked sites. 

\subsection{Predicting ligand-specific binding sites}
Recently, deep learning models that predict protein-ligand complex structures, given a protein-ligand pair, have been developed (\cite{deepdock}, \cite{geom_DL_binding_conformation}, \cite{equibind}). In principle, these models can be used to find ligand-specific binding sites. Therefore, one may argue that the BSP models are strictly less useful than these models, since their predictions on binding sites do not take into account the partner ligands. However, we argue that they are still useful in their own right. Firstly, for many applications, predicting ligand-agnostic binding sites is enough, and even desirable. For example, a typical docking experiment requires a binding pocket location as a prerequisite, and docks all molecules in a virtual library to the pocket. To predict binding pockets using the models that do consider the partner ligands, preparing appropriate ligands may add additional complexity to the problem. This is similar to the problem of preparing appropriate \textit{probe} molecules in the previously mentioned \textit{prob-based methods}. 
Secondly, the performances of the methods that predict protein-ligand complex structures are not satisfactory at this point. For example, EquiBind (\cite{equibind}) scored median and mean ligand rmsd of $6.2\AA$ and $8.2\AA$, which suggests that the model is not accurate enough to correctly identify binding residues. Therefore, we might want to focus on the easier and well-studied BSP problem. 

\section{Problem Definition}\label{section:problem_definition}

Thus far, the BSP task has not been addressed explicitly under a common definition across different literatures, even though they used different model compositions. For example, while Deeppocket is comprised of separate ``detection" and ``segmentation" models, Kalasanty only uses single segmentation model, whose output is post-processed by a clustering algorithm.

In order to fairly assess different BSP models, it is necessary to envisage an unified definition of the BSP task. To be more specific, we will formally establish standards on the input and output. All baseline models can be regarded as following the standard, which will be explained in the experiment section.

Moreover, we will also explain the decomposition of the task into sub-tasks (including BSD and BRI), which is employed in our method and Deeppocket.

\subsection{The BSP task}

BSP is the task of identifying the ligand binding sites in a given protein. 
In the task, we are given as input: a protein structure $P$ and the number of binding sites $n$. We assume that there are known structures of ligands $l_i$ ($i=1,\cdots,n$) that correspond to the binding sites. 

The goal of the task is to predict an unordered set of $n$ binding sites of $P$ where the ligands $l_1,\cdots,l_n$ bind.

A \textit{predicted binding site} is of the form $(\hat{c}_i, \hat{R}_i)$, where $\hat{c}_i\in\mathbb{R}^3$ is the \textit{binding site center} and $\hat{R}_i\subset \set{1,\cdots,size(P)}$ is the set of indices of \textit{binding residues}. For example, an ideal prediction $\set{(\hat{c}_1,\hat{R}_1),\cdots,(\hat{c}_n,\hat{R}_n)}$ is such that 
\begin{itemize}
\item $\hat{c}_i$ is close (e.g. within the radius threshold $4\AA$) to $l_i$
\item $\hat{R}_i$ is the set of indices of residues close (e.g. within the radius threshold $4\AA$) to $l_i$
\end{itemize}

The methods that we used to evaluate such predictions will be explained in \ref{subsection:evaluation}. 

\subsection{Decomposition into sub-tasks}\label{subsection:problem_division}

Our method divides the BSP task into sub-tasks (1) candidate generation, (2) Binding Site Detection (BSD) and (3) Binding Residue Identification (BRI), each corresponding to a dedicated module (this is similar to TM-SITE (\cite{tmsite_ssite}) and Deeppocket (\cite{Deeppocket})). To be more specific, let $(P,n)$ be an input to perform BSP on. First, the \textit{candidate generation module} takes the protein structure $P$ as an input then generates the candidate binding site centers $\hat{c}'_1, \hat{c}'_2, ..., \hat{c}'_m\in\mathbb{R}^3$, where typically $m \gg n$. Next, the \textit{BSD module} takes $(P, \hat{c}_i)$ ($1\le i\le m$) as the inputs and outputs the predicted \textit{druggability} of $\hat{c}'_i$ in $P$. The druggability scores are then used to rank the candidate centers, the top $n$ of which form a filtered list  $\hat{c}_1,\cdots,\hat{c}_n$ of candidate centers. Lastly, for each $1\le i\le n$, the \textit{BRI module} takes as input $(P, \hat{c}_i)$, and outputs $\hat{R}_i$, that is the set of binding residues within the binding site. The resulting set $\set{(\hat{c}_1, \hat{R}_1),\cdots,(\hat{c}_n,\hat{R}_n)}$ becomes the final output of the model.

\section{Key Components of Our Method}\label{section:components}

\begin{figure}
\centering
\includegraphics[width=18cm]{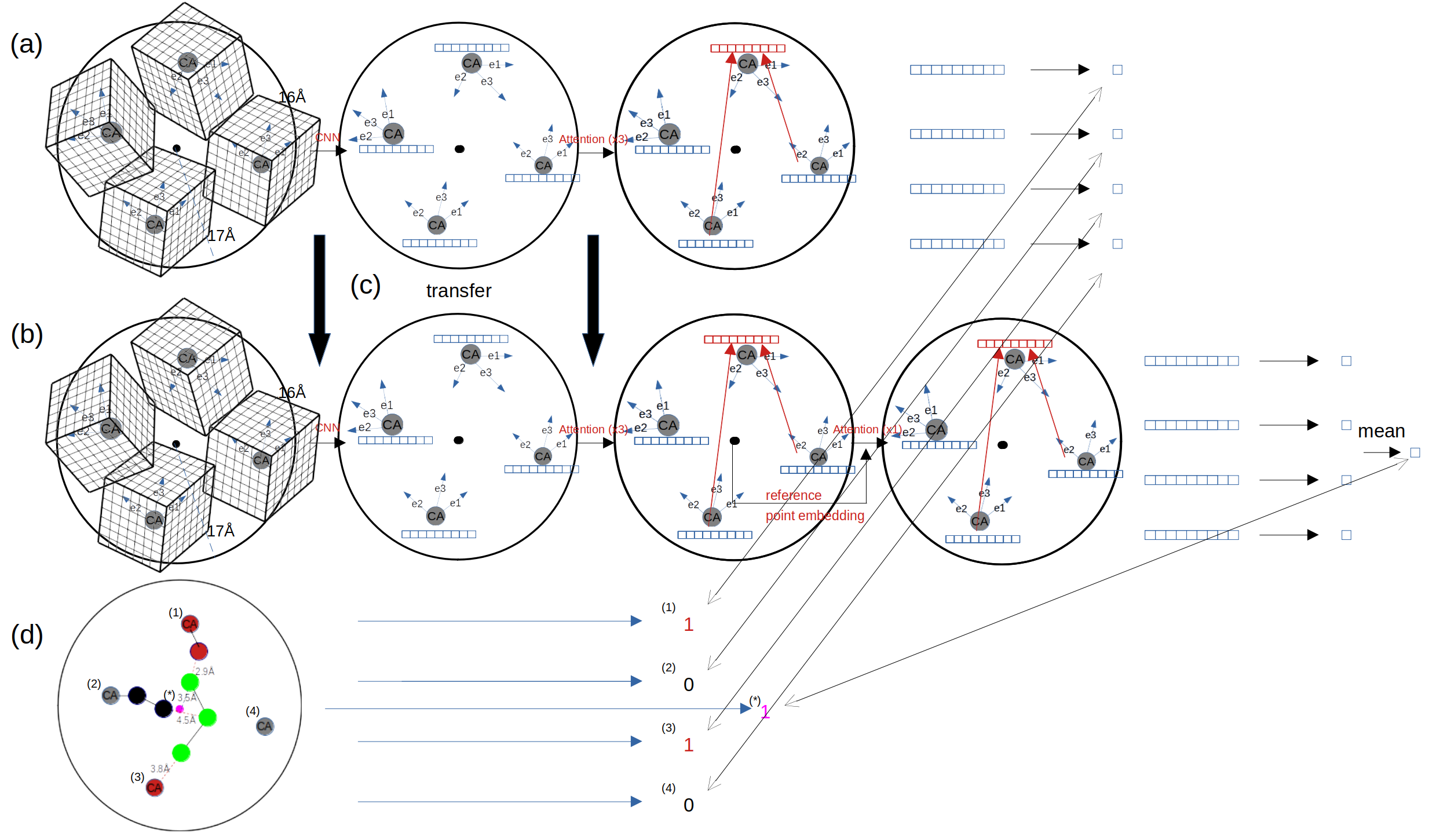} %
\caption{
(a) Our BRI module.
(b) Our BSD module.
(c) Our BSD module is trained in two stages, where in the second stage, the parameters of the shared parts are initialized from the result of training a BRI module in the first stage. 
(d) Ground truth generation. A candidate site is a true binding site if the center is within $4\AA$ from a ligand atom (i.e. $DCA < 4\AA$). A residue is a binding site residue if it is within $4\AA$ from a ligand atom.
}
\label{model_figure}%
\end{figure}

\begin{figure}
\centering
\includegraphics[width=18cm]{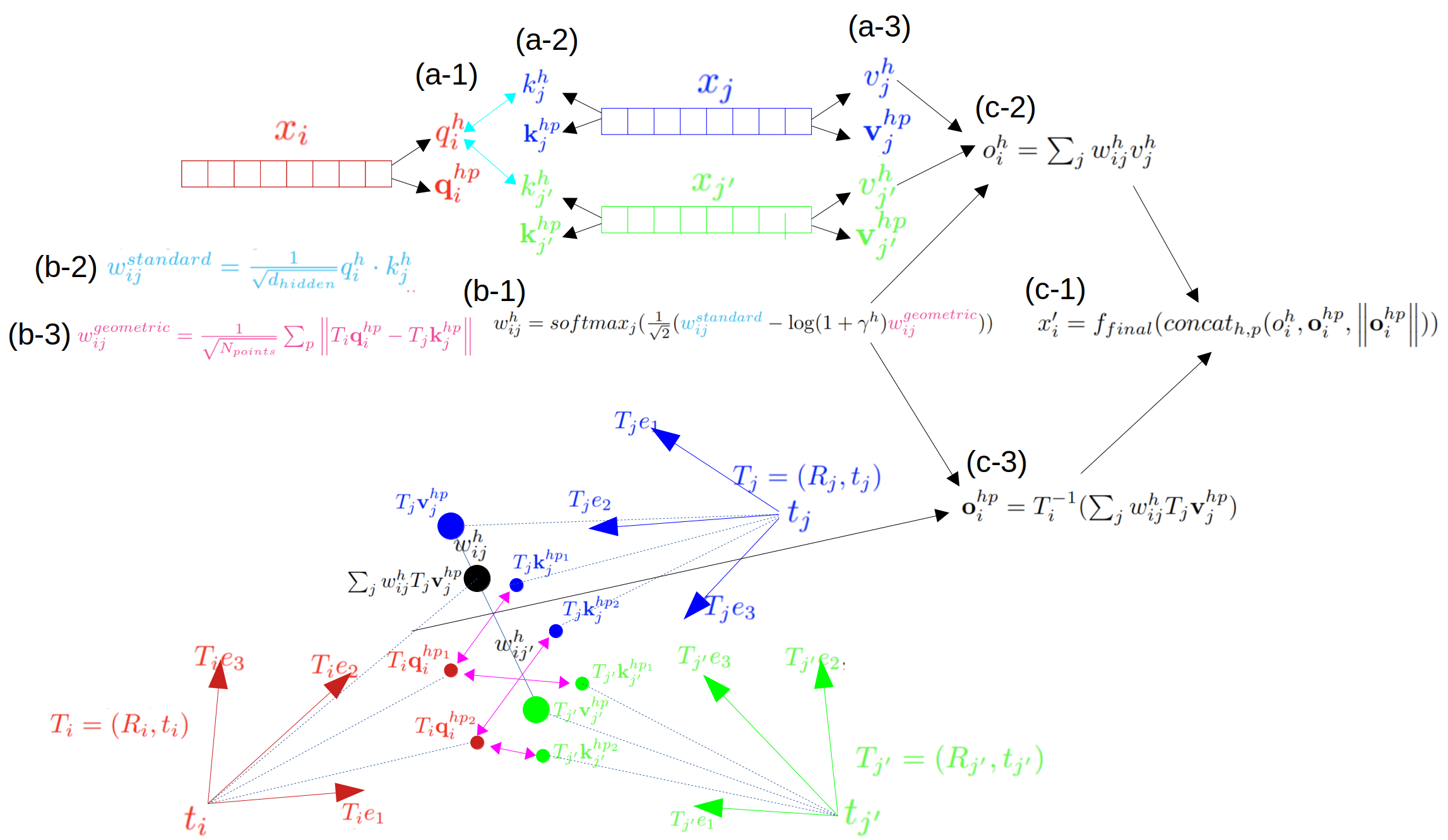} %
\caption{
The figure illustrates, given hidden vectors $\set{x_i}_{i=1}^n$ and geometric information $\set{(R_i,t_i)}_{i=1}^n$, how our attention layer produces updated hidden vectors $\set{x_i'}_{i=1}^n$.
(a) Obtaining query, key and value vectors required for the computation. 
(a-1) Standard ($q_i^h$) and geometric ($\textbf{q}_i^{hp}$) query vectors.
(a-2) Standard ($k_i^h$) and geometric ($\textbf{k}_i^{hp}$) key vectors.
(a-3) Standard ($v_i^h$) and geometric ($\textbf{v}_i^{hp}$) value vectors.
(b) Obtaining attention weights. 
(b-1) The attention weights are based on the standard terms ($w_{ij}^{standard}$) and the geometric terms $w_{ij}^{geometric}$.
(b-2) The standard terms are based on the match (measured by inner product) between the standard query and key vectors. 
(b-3) The geometric terms are based on the match (measured by the distance between their global coordinates) between the geometric query and key vectors. 
(c-1) The updated hidden vectors are calculated based on the aggregated standard value vectors ($o_i^h$) and the aggregated global coordinates of the value vectors ($\textbf{o}_i^{hp}$).
(c-2) Aggregating the standard value vectors.
(c-3) Aggregating the global coordinates of the geometric value vectors. 
}
\label{att_figure}%
\end{figure}

This section briefly illustrates the key components of our method, which will be explained in more details in \secref{section:methods}. These include the details of the modules (candidate generation, BSD and BRI) as well as other aspects independent to the model architecture. In particular, the latter includes the transfer learning and the homology-based augmentation method.

\subsection{The candidate generation module}

To generate the binding site candidate centers, we use an external software Fpocket (\cite{Fpocket}). Given a protein structure, Fpocket finds sets of heavy atoms $\hat{S}_1,\cdots, \hat{S}_m$, each corresponding to a region geometrically likely to be a binding pocket. Then, we find the candidate centers $\hat{c}'_i$ ($i=1,\cdots,m$) by taking the center of the mass of the atoms in $\hat{S}_i$. 

We chose Fpocket as the candidate generation method because it achieves a sufficiently high recall rate (96.4\% on scPDB v.2017, according to \cite{Deeppocket}). This means that, for a given protein and its binding site, it is likely that at least one of the generated candidates corresponds to the binding site. Then, provided that the BSD module ranks the candidates properly, the top-$n$ candidates may approximate the true binding site centers with a high accuracy. 

\subsection{The BSD module}\label{subsection:bsd_module}
The BSD module takes as input the protein structure and a candidate binding site center $\hat{c}'$, and outputs the predicted druggability at $\hat{c}'$.

In doing so, it featurizes the surroundings of $\hat{c}'$ into a set of per-residue 3D grids, and processes the grids through a neural network to produce the output. Here, each grid in the set corresponds to a residue close enough to $\hat{c}'$ (distance threshold $17\AA$), and encodes the local environment of the residue.   

%The grid featurization process is composed of the following steps:
%\begin{enumerate}
%\item Collect residues whose alpha carbons are within a radius threshold ($17\AA$) from $\hat{c}'$. 
%\item For each residue, find a system of orthogonal axes.
%\item Lay cubical grids centered at the residues and aligned with the axes. 
%\item Featurize the grid voxels according to the atoms in them. \end{enumerate} 
%The mathematical detail of these steps will be explained in \ref{section:methods}. 

The neural network of the BSD module is composed of (1) a residue-local feature extracting unit which runs in parallel for each grid (2) an aggregation unit which globally aggregates the local features (3) the reduction unit which maps the aggregated feature to single scalar quantity. The feature extracting unit is a 3D CNN model, and the aggregation unit is composed of several geometric self-attention layers. The reduction part is composed of a point-wise feed-forward layer and a mean-reduction operation. 

\subsection{The BRI module}\label{subsection:bri_module}

The BRI module takes in the protein structure and a putative binding site $\hat{c}$ as inputs and outputs the set of predicted binding residue indices. 

The BRI module shares the residue-local feature extraction and global aggregation units with the BSD module. To be more specific, the BRI module shares the following units with the BSD module: (1) the CNN feature extractor and (2) the stack of geometric attention layers up to the penultimate one in the BSD module. However, the remaining part of the BRI module is only comprised of a point-wise feed-forward layer without a mean-reduction operation. Hence, the outputs of the last layer are used to determine (with a threshold value) whether the corresponding residues are binding site residues or not. 

\subsection{Transfer learning}\label{subsection:transfer_learning}
Transfer learning can be applied between the BSD and BRI sub-tasks thanks to the shared architectures between the BSD and BRI modules. More specifically, we initialize the weights of the BSD module's shared parameters with the weights obtained from BRI module's training. The rationale behind this procedure is the following intuition: the protein’s binding site can be determined based on the patterns of the binding residues. Under this rationale, we hypothesize that a well-performing BRI module will learn useful features that can transfer well to the BSD task.

In addition, the transfer learning allows BSD module to leverage relatively more abundant labels present in BRI dataset. While there is one label per binding site for the BSD task, there are a multiple number of binding residues per binding site. Thus, it is desirable to exploit such abundance in labels from the BRI task  for BSD task via the transfer learning. 

%(explained in the Experiments section in more detail)`

\subsection{The geometric self-attention layers}\label{geometric_attention}

As mentioned previously, we adopted the geometric self-attention layers to globally aggregate local features obtained by the CNN. First, let $\set{x_i}{i=1}^n$ denote a sequence of hidden vectors that is either the initial local features computed by the CNN or the output of the preceding attention layer. Then a geometric self-attention layer $f_{att}$ transforms $\set{x_i}{i=1}^n$ into another sequence of hidden vectors $\set{x_i'}{i=1}^n$ based on the protein structure. In our method, we represent the protein structure as \textit{local frames} (\cite{alphafold}) $\set{T_i}_{i=1}^n=\set{(R_i,t_i)}_{i=1}^n$ associated to the residues, where $R_i$ is the \textit{residue orientation} and $t_i$ is the \textit{residue center}. 
Then the geometric self-attention layer $f_{att}$ takes the following form:
\begin{equation}\label{geom_att_schema}
\set{x_i'}_{i=1}^n= f_{att}\left(\set{x_i}_{i=1}^n, \set{T_i}_{i=1}^n\right)
\end{equation}

The geometric attention mechanism has several advantages over the modules from previous works used for the feature aggregation---the traditional clustering algorithms and the CNN models. First, unlike the Neural Network based methods (such as our), the clustering algorithms are less flexible in terms of the number of adjustable parameters and they do not fit their parameters based on the gradients. As for the CNN models, the attention layers are arguably more effective in terms of modeling the long-distance dependency. While an attention layer can emulate arbitrarily distant interaction by a single step of operation, convolution layers require several steps to do so due to their local nature. This \textit{long-term dependency} may lead to sub-optimal learning outcomes (\cite{grad_flow_rnn}).

For the attention mechanism, we used a modified version of \textit{Invariant Point Attention} from Alphafold (\cite{alphafold}). The essential ideas of its computation, compared to the standard attention mechanism (\cite{vaswani}), are that (1) it uses not only the standard \textit{query}, \textit{key} and \textit{value} vectors but also geometric ones, (2) it calculates the attention weights based on not only the inner products of the standard query and key vectors but also the distances between the geometric query and key vectors, and (3) its output is determined by not only the aggregated standard value vectors but also the aggregated geometric value vectors. More detailed descriptions are provided in \figref{att_figure} and the Methods section. 
The effect of using the geometric vectors in the attention mechanism is discussed as a part of our ablation study. 

%The simplification in our model compared to the original Invariant Point Attention was that it omitted a term in the calculation of attention weights that was irrelevant in our problem/model. This was because our problem/model didn't have a correspondence for the ``pair representation" of Alphafold, which encoded relations between pairs of residues as computed from the preceding parts of the model. The use of our modified IPA of AlphaFold indeed led to performance improvement.

\subsection{Satisfaction of SE(3)-invariance}\label{subsection:se3_invariance}

A function is \textit{SE(3)-invariant} if its output remains unchanged when SE(3) transformations (translations, rotations or compositions of them) are applied to the input. In our context, when $\set{v_i}_{i=1}^N$ are the coordinates of the protein atoms, $\set{f_i}_{i=1}^N$ are the feature vectors of the protein atoms, and $\set{T_i}_{i=1}^n$ are the local frames related to the protein residues, a function $f(\set{v_i}_{i=1}^N, \set{f_i}_{i=1}^N \set{T_i}_{i=1}^n)$ is SE(3)-invariant if, for any SE(3) transformation $T$, we have 
\begin{equation*}
f(\set{Tv_i}_{i=1}^N, \set{f_i}_{i=1}^N, \set{TT_i}_{i=1}^n)=f(\set{v_i}_{i=1}^N, \set{f_i}_{i=1}^N, \set{T_i}_{i=1}^n)
\end{equation*}

The SE(3)-invariance is a desired property for the structure-based BSP models. This is because the binding site information should remain unchanged regardless of the reference frame. To reflect this, we are injecting this inductive bias via incorporating the SE(3)-invariance property into our model and hence achieve robustness in our model’s prediction (\cite{se3_transformers}).

Two strategies were employed to achieve SE(3)-invariance in our model. First, we adopted SE(3)-invariant attention mechanism at the outset. To be more specific, we required the attention layer in  \eqref{geom_att_schema} to be invariant to any SE(3) transformation $T=(R,t)$, that is:
\begin{equation}\label{att_se3_invariance}
f_{att}(\set{x_i}_{i=1}^n, \set{TT_i}_{i=1}^n) = f_{att}(\set{x_i}_{i=1}^n, \set{T_i}_{i=1}^n). 
\end{equation} 
The details of the SE(3)-invariant attention mechanism will be explained in the Methods section. Then, the other strategy employed to meet the SE(3)-invariance was altering the grid-based residue featurization process. When constructing the grids, we did not use an arbitrary (xyz-) axes, but rather used axes aligned with respect to the orientations of the residues. This will be also explained in more detail in the Methods section.

\subsection{Augmentation strategies}\label{augmentation_strategies} 
\textit{Data augmentation} is one of the most essential elements comprising the modern Deep Learning.  It plays a crucial role in (1) alleviating the problem of overfitting which occurs commonly due to the lack of data and (2) improving the trained model’s generalization capability (\cite{augmentation_survey}). In essence, the data augmentation enlarges the effective size of the training set by applying various transformations to the inputs.

Previous Deep Learning methods for BSP (\cite{DeepSite}, \cite{Deeppocket}, \cite{Kalasanty}) employed various augmentations. These augmentation methods were mainly based on applying transformations in the SE(3) class (rotations, translations and compositions of them) to the inputs of the CNN models.

In the absence of data augmentation, our model suffered from a clear pattern of over-fitting (\figref{overfitting_curves}). We believe that this is due to the lack of diversity in the training dataset as compared to the large complexity of the model.

However, the existing augmentation methods based on SE(3) transformations have no effect on the training dynamics when used in conjunction with our SE(3)-invariant model. From this aspect, we devised two novel augmentation methods: (1) a method based on a class of geometric transformations and (2) a method based on the homologous proteins.

Firstly, the augmentation based on the geometric transformation introduces the \textit{random perturbations} to the residue orientations.   This differs to the usual ``random rotation" in that it permits only rotations of small magnitude. We applied independent random perturbations on the residue orientations once then used the modified residue orientations as an input to every attention layer. The rationale behind applying such random perturbation is to promote diversity in geometric information while not completely forgetting the original residue orientation.
%while not breaking the SE(3)-invariance of the model. 

Next, we employed an augmentation scheme based on the protein homology.  This fundamentally differs from the usual augmentation methods, in that it relies on an unlabelled external database of unbound protein structures to pre-compute the transformations.  Essentially, we first align the protein sequences in the training set (referred to as a \textit{internal protein sequence}) with the protein sequences from the external database (referred to as a \textit{external protein sequence}). Then, we assign the ground-truth binding site and binding residue labels of the internal protein sequence to the aligned external protein sequence as its labels. Finally, we augment our training dataset with these aligned external protein sequences and their labels assigned. The extended dataset may contain some label noises since the assignment of labels based on the homology relations is subject to inaccuracies to some extent. In spite of this problem, our augmentation scheme enhances diversity of input data compared to the previous augmentation schemes. This is because the extended dataset may include protein species absent in the training dataset.

\section{Experiments}\label{section:experiments}

\subsection{Datasets}\label{subsection:datasets}
We used scPDB v.2017 (\cite{scPDB}) as a main dataset for the training and the validation. In addition, we used three other datasets for the tests: COACH420 (\cite{p2rank}), HOLO4K (\cite{p2rank}) and CHEN (\cite{chen_dataset}). To be more specific, we used the training subset of scPDB dataset provided by \cite{DeepSurf} for 5-fold cross-validation. This subset excludes proteins which have sequence similarity higher than 90\% to the proteins in one of the external test datasets.
We used the remaining part of the scPDB as a test dataset.
Thus, the test datasets were comprised of the scPDB test set and the external test dataests --- COACH420, HOLO4k and CHEN.
The CHEN dataset had holo and apo subsets. Thus in the tests using the apo subset, we obtained the ground-truth binding sites from the structural alignments with the corresponding holo structures. More specifically, the ligands of the holo structures were superimposed onto the apo structures according to the structural alignments. The characteristics of each test dataset and the details of the structural alignments are described in the Supplementary Information.

\subsection{Baseline Methods}

We compared our method to the previous state-of-the-art Deep Learning methods, which are based on CNN: Deeppocket (\cite{Deeppocket}), Kalasanty (\cite{Kalasanty}) and DeepSurf (\cite{DeepSurf}). All these methods are briefly explained in \ref{subsection:dl_based-related_works}. For Deeppocket and DeepSurf, we have trained their parameters from scratch according to our dataset splits. However for Kalasanty, we used the parameters released by the authors due to the high computational costs to train. It is important to note that the parameters of Kalasanty were trained on the entire scPDB v.2017 dataset. To be specific, the training data of Kalasanty may have included data whose protein sequences are similar (similarity above 90\%) with those in the test dataset. Thus, Kalasanty method have an advantage in terms of the coverage of the training dataset compared to the other methods when they are evaluated on the external test datasets.
  
\subsection{Ablated Methods}

We conducted an ablation study to assess the effectiveness of each component of our proposed method. We considered the omission of the following components:
\begin{itemize}
\item The geometric information used in the self-attention layers
\item The use of local features extracted by the CNN 
\item The grid alignment process adopted to achieve SE(3)-invariance in our model
\item The transfer learning from the BRI to the BSD module
\item The data augmentation methods as a whole 
\item Each of the data augmentation methods 
\end{itemize} 

We used the most basic form of transformers, that is BERT (\cite{vaswani}), as a geometric information omitted version of self-attention layers to conduct the ablation study on `the geometric information used in self-attention layers'.

In the ablation study of `the use of local features extracted by the CNN', we removed the CNN component from our model. To be more specific, the hidden vectors for the attention layers are directly obtained from the One-hot-encoding layer for the amino acid types followed by the token embedding layer. To compensate for the loss of model complexity, we added two more attention layers to the default configuration of BRI and BSD model architecture.

In the ablation study on the grid alignment process, we removed the alignment process but used the standard random rotation augmentation for a fair comparison. By doing so, we attempted to argue the efficacy of our method as compared to the common practice, rather than the case where no additional technique is applied. 

\subsection{Evaluation}\label{subsection:evaluation}

% Also I think it is good to provide an introductory paragraph for evaluation metric. This paragraph should introduce that we need evaluation metric for each task i.e. "BSD" and "BRI" and give some overview picture of evaluation metric.
% => 반영함. (첫 2개 paragraph) 

In \ref{section:problem_definition}, we formalized the inputs and outputs of the BSP methods. According to the formalism, a BSP method (1) takes as inputs a protein structure and the number of binding sites ($n$), and (2) outputs $n$ predictions $(\hat{c}_1,\hat{R}_1),\cdots,(\hat{c}_n, \hat{R}_n)$ with $\hat{c}_i$ and $\hat{R}_i$ being the binding site centers and the binding residue indices, respectively.
Now that we have formalized the problem, we need to establish an evaluation scheme. This requires (1) interpreting each method's input and output as per the problem definition and (2) defining the evaluation metrics in terms of the input and output formats specified in the problem definition.

\subsubsection{How the baseline methods fit into our problem definition}

The definitions of $\hat{c}_1,\cdots, \hat{c}_n$ for the baseline methods are mostly natural, and derived directly from the original papers. All methods produce a ranked list of predictions, thus we limit them to produce only the top-$n$ outputs. Also, they compute the centers of predicted binding sites in their evaluation, so we can compute $\hat{c}_i$ as they prescribed.

However, not all baseline models output the predicted binding sites at the residue level. Thus, it is necessary to map their outputs to sets of residues $\hat{R}_1,\cdots, \hat{R}_n$. For example in the Deeppocket (\cite{Deeppocket}), the authors used the distance threshold $2.5\AA$ (performed best in their validation set) to determine the binding residues from the segmented voxels; therefore, we followed the same procedure. For Kalasanty (\cite{Kalasanty}) and DeepSurf (\cite{DeepSurf}), the authors introduced a method to convert their predictions to the atom-level predictions (which was implemented in their code); therefore, we regarded the residues having at least one such predicted binding atoms as the predicted binding residues.

\subsubsection{The evaluation metrics}\label{subsubsection:evaluation_metrics}

We use three evaluation metrics: (1) \textit{the success rate for detection} (success rate) (2) \textit{the average IOU of binding residues with respect to the closest ligands} (IOU) (3) \textit{the average IOU of the binding residues with respect to the successfully detected ligands} (conditional IOU). The metrics evaluate different combinations of BSD and BRI performances. The success rate metric evaluates the BSD performance, the IOU metric evaluates the BRI performance but it is also influenced by the BSD performance, and the conditional I metric aims to evaluate the BRI performance alone. We give additional details with regards to how these evaluation metrics compare to their counterpart metrics introduced in the previous literatures in Supplementary Information.

In order to provide a formal definition of each metric, we shall adopt following notations: 
\begin{itemize}
\item $n^{(i)}$ is the number of ground-truth ligands bound to the $i$-th protein. 
\item $\set{l^{(i)}_{1},\cdots, l^{(i)}_{n^{(i)}}}$ is the set of ground-truth ligands bound to the $i$-th protein. \item $\set{(c^{(i)}_1,BR^{(i)}_1)\cdots, (c^{(i)}_{n^{(i)}}, BR^{(i)}_{n^{(i)}})}$ is the set of predictions of the method to evaluate.
\item $\set{TBR^{(i)}_1,\cdots,TBR^{(i)}_{n^{(i)}}}$ is the set of \textit{true binding residue indices}, where $TBR^{(i)}_j$ is defined to be the set of residues in the $i$-th protein that is within $4\AA$ from $l^{(i)}_j$
\end{itemize}

The success rate metric measures the correspondence between the predicted binding site centers $\set{c^{(i)}_1,\cdots,c^{(i)}_{n^{(i)}}}$ and the positions of the ground-truth ligands $\set{l^{(i)}_1,\cdots,l^{(i)}_{n^{(i)}}}$. For each $i$, we compute the F1 score (the harmonic mean of precision and recall) based on the definition of \textit{detection}. Specifically, we define that $c^{(i)}_j$ is a \textit{correct detection} of $l^{(i)}_k$ when $c^{(i)}_j$ is within $4\AA$ (a threshold commonly used in the literature e.g. \cite{Deeppocket} and \cite{DeepSurf}) from any ligand of $l^{(i)}_k$. In other words, we define that detection is correctly performed when the Distance from Center to Atom (DCA) is $<4\AA$. Then, the F1 scores are weighted-averaged (weighted by $n^{(i)}$) over the proteins. In summary, we obtain this metric as

\begin{equation*}
\left (\sum_{i}n^{(i)}\cdot\frac{2}{\frac{1}{P^{(i)}}+\frac{1}{R^{(i)}}} \right ) \bigg / \left (\sum_{i}n^{(i)} \right)
\end{equation*},
where $P^{(i)}$ is the \textit{precision} defined as follows:
\begin{equation*}
    P^{(i)} = \frac{\#\set{1\le j\le n^{(i)}:c^{(i)}_j\text{ detects one of }l^{(i)}_1,\cdots l^{(i)}_{n^{(i)}}}} {n^{(i)}}
\end{equation*}
 and $R^{(i)}$ is the \textit{recall} defined as follows:
\begin{equation*}
 R^{(i)} = \frac{\#\set{1\le k\le n^{(i)}:l^{(i)}_k\text{ is detected by one of }c^{(i)}_1,\cdots,c^{(i)}_{n^{(i)}}}}{ n^{(i)}}
\end{equation*}
 
This is a BSD metric since this involves only the predicted binding site centers, not the predicted binding residues. 

The IOU metric compares the predicted binding residues $BR^{(i)}_j$ with the true binding residues $TBR^{(i)}_{\phi^{(i)}(j)}$ of the ligand $l^{(i)}_{\phi^{(i)}(j)}$ closest to the predicted binding site center $c^{(i)}_j$. Here, the index $\phi^{(i)}(j)$ of the closest ligand is defined as 
\begin{equation*}
    \phi^{(i)}(j) = \argmin_{k=1}^{n^{(i)}}DCA(c^{(i)}_j, l^{(i)}_k)
\end{equation*}
The comparison is performed in terms of Intersection Over Union (IOU), and the quantity is averaged over all pairs of $(i, j)$. In summary, we obtain the second metric as 
\begin{equation*}
\left(\sum_{i}\sum_{j=1}^{n^{(i)}}\frac{\#(BR^{(i)}_j\cap TBR^{(i)}_{\phi^{(i)}(j)})}{\#(BR^{(i)}\cup TBR^{(i)}_{\phi^{(i)}(j)})}\right) \bigg/ \left(\sum_{i}n_i\right)
\end{equation*}

Although this is essentially a BRI metric, it also depends on the BSD performance due to the definition of $\phi^{(i)}(j)$. In particular, if the predicted center $c^{(i)}(j)$ is far from any ligand, the set of predicted binding residues $\hat{R}_j^{(i)}$ does not contribute to the metric.

The conditional IOU metric is almost the same as the IOU metric, but it aims to eliminate the previously mentioned problem that the BRI performance is bound by the BSD performance. It does so by focusing on the case that the predicted binding sites is close to at least one ligand. In summary, we obtain the metric as 
\begin{equation*}
\left (\sum_{i}\sum_{j\in S^{(i)}}\frac{\#(BR^{(i)}_j\cap TBR^{(i)}_{\phi^{(i)}(j)})}{\#(BR^{(i)}\cup TBR^{(i)}_{\phi^{(i)}(j)})}\right ) \bigg / \left(\sum_{i}\# S^{(i)} \right)
\end{equation*}
, where 
\begin{equation*}
    S^{(i)} = \set{j=1,\cdots,n^{(i)}:DCA(c^{(i)}_j,l^{(i)}_{\phi^{(i)}(j)})<4\AA}
\end{equation*}
This is similar to metrics used in Deeppocket (\cite{Deeppocket}) and DeepSurf (\cite{DeepSurf}) to evaluate models' binding site \textit{segmentation} capability conditional on the successful location of the binding sites. 

%Although this metric aims to measure the models' segmentation capability alone independent of the detection capability, \red{(re-write this)}
%Note that it may be unfair to directly compare differ methods based on this metric, since the methods might result in different sets $S^{(i)}$ on which to compute the average IOU. Therefore, we record this metric only as a reference. For more objective comparisons, one has to refer to the second metric instead. 

\subsection{Training the BSD Module}\label{subsection:BSD_training}

To implement the transfer learning described in \ref{subsection:transfer_learning}, the BSD training consists of two stages. 
The first stage is pre-training the part of the BSD module's architecture shared by the BRI module, as depicted in \figref{model_figure}. In doing so, we attach the unshared part of the BRI module architecture on top of the shared part in the BSD module then train the combined model for the BRI task. The second stage is fine-tuning the entire original BSD module for the BSD task. In the second stage, to promote smoother transfer learning, we freeze the parameters of the parts trained in the first stage up to certain steps of gradient descent. 
%We adopt this two-stage procedure to maximally utilize the information provided by the training dataset. This is because using only the BSD labels may lead to potential under-utilization of the more detailed information in the resolution of residues. Also, this can be viewed as reinforcing the previously noted intention that ``the attention layers globally aggregate the local information". This is because the unshared parts in the BSD module make predictions on the binding site label based on the last hidden layer activations of a trained BRI module. 

In both stages, we use balanced sampling of binding site candidates of positive and negative labels. This is because, among the binding site candidates predicted by Fpocket (On average 33 per protein in scPDB), only few of them are actual binding sites (typically only one). Training without such balanced sampling may render the trained model biased toward the out-numbered label. (\cite{class_imbalance_survey})
 
In addition, in the first stage, we resolve the similar problem of unbalanced residue labels by using a weighted loss function. This loss function consists of a weighted sum of terms from different residues, where the binding and non-binding residues attain the following weights:
\begin{equation*}
    w_{pos}=\frac{1}{2n_{pos}},\quad w_{neg}=\frac{1}{2n_{neg}}
\end{equation*}
, where $n_{pos}$ and $n_{neg}$ are the number of binding and non-binding residues respectively.

%First stage - weighted binary cross entropy loss

%For a batch, the resulting weighted sums are mean-reduced to obtain the loss.  
%We use minibatch size 16, and peak learning rate 1e-4. We use AdamW optimizer with $\beta_1=0.9$, $\beta_2=0.999$ and weight decay of rate 0.01 and the cosine learning rate scheduler (the most simple form of cosine annealing \cite{cosine_annealing}) with 6000 warmup steps and 60000 total steps.   

%In the second stage, we use mean-reduced binary cross entropy loss, minibatch size 16, peak learning rate 1e-6, AdamW optimizer with $\beta_1=0.9$, $\beta_2=0.999$ and weight decay of rate 0.01, and the cosine learning rate scheduler with 4000 warmup steps and 20000 total steps. We freeze the model parameters of the parts shared by the BRI module architecture until 8000 steps, to promote smoother fine-tuning of the parameters.   

\subsection{Training the BRI module}\label{subsection:BRI_training}

To train the BRI module, we use only the positive binding site candidates. (We do the same to train Deeppocket) This is because, in our intended usage, the BRI module operates on the binding sites detected by the BSD module. Note that this intention is reflected in the evaluation metric \textit{average IOU
of binding residues against the closest ligands} as well. All the settings of the first stage of BSD training were maintained, except the balanced sampling of the binding site candidates.

\subsection{Data Augmentations}\label{applying_augmentations}
We apply the data augmentation methods described in \ref{augmentation_strategies} in every training scenario (pre-training the BSD module, fine-tuning the BSD module and training the BRI module), except when they are omitted as a part of the ablation study. Applying the random perturbation means transforming the inputs to the geometric attention layers once for each data sample. Applying the homology-based ``augmentation" means adding to the original loss (originated from the original dataset) an auxiliary loss calculated in the same way but originated from the ``extended dataset".

\subsection{Experiment Results}

\begin{table}
	\tiny
	\centering
	\caption{(BSD metric) F1 Success rate for detection. The mean and standard deviation are calculated based on the metric values for 5 different cross-validation folds.}
	\label{t1}
	\begin{tabular}{|c|c|c|c|c|c|}
		\noalign{\smallskip}\noalign{\smallskip}\hline & scPDB(held-out) & COACH420 & HOLO4k & CHEN-holo & CHEN-apo \\
		\hline 
		DeepSurf & $62.4\pm 1.3$ & $43.6\pm 1.3$ & $59.7\pm 1.4$ & $24.5\pm 1.4$ & $22.3\pm 1.2$ \\
		Kalasanty & $70.0\pm 0.0$ & $50.8\pm 0.0$ & $44.9\pm 0.0$ & $28.5\pm 0.0$ & $27.1\pm 0.0$ \\
		DeepPocket & $67.9\pm 0.4$ & $55.7\pm 0.7$ & $72.2\pm 0.2$ & $\bm{42.4\pm 0.3}$ & $34.5\pm 1.2$ \\

		\hline
		Ours  & $\bm{70.1\pm 0.4}$ & $59.1\pm 0.3$ & $\bm{77.0\pm 0.6}$ & $41.2\pm 1.2$ & $36.5\pm 0.2$ \\
		Ours(BERT) & $65.6\pm 1.2$ & $54.8\pm 1.2$ & $70.8\pm 1.1$ & $38.2\pm 1.2$ & $32.8\pm 0.7$ \\
		Ours(no CNN) & $69.2\pm 0.7$ & $57.7\pm 0.3$ & $75.1\pm 0.2$ & $41.7\pm 1.0$ & $\bm{36.7\pm 1.0}$ \\
		Ours(no alignment)  & $69.0\pm 0.6$ & $\bm{59.4\pm 0.7}$ & $75.6\pm 0.7$ & $41.0\pm 0.6$ & $35.4\pm 0.6$ \\
		Ours(no transfer) & $61.9\pm 0.3$ & $53.9\pm 0.3$ & $70.0\pm 0.5$ & $39.8\pm 0.5$ & $35.1\pm 0.6$ \\
		Ours(no augmentation) & $64.2\pm 2.0$ & $55.8\pm 1.4$ & $71.3\pm 1.5$ & $39.1\pm 1.7$ & $34.3\pm 0.5$ \\
		Ours(no homology) & $68.3\pm 1.1$ & $57.3\pm 0.9$ & $75.1\pm 1.1$ & $39.9\pm 0.5$ & $34.6\pm 0.6$ \\
		Ours(no perturbation) & $70.0\pm 0.5$ & $58.9\pm 0.7$ & $76.2\pm 0.3$ & $41.8\pm 0.7$ & $36.3\pm 1.1$ \\
		\hline
	\end{tabular}
\end{table}

\begin{table}
	\tiny
	\centering
	\caption{(BSD + BRI metric) Average IOU of binding residues against the closest ligands. The mean and standard deviation are calculated based on the metric values for 5 different cross-validation folds.}
	\label{t2}
	\begin{tabular}{|c|c|c|c|c|c|}
		\noalign{\smallskip}\noalign{\smallskip}\hline & scPDB(held-out) & COACH420 & HOLO4k & CHEN-holo & CHEN-apo \\
		\hline 
		DeepSurf & $0.288\pm 0.007$ & $0.194\pm 0.006$ & $0.207\pm 0.005$ & $0.104\pm 0.003$ & $0.085\pm 0.005$ \\
		Kalasanty & $0.260\pm 0.000$ & $0.183\pm 0.000$ & $0.146\pm 0.000$ & $0.101\pm 0.000$ & $0.092\pm 0.000$ \\
		Deeppocket & $0.440\pm 0.002$ & $0.313\pm 0.003$ & $0.277\pm 0.03$ & $0.190\pm 0.005$ & $0.186\pm 0.003$ \\
		\hline
		Ours & $\bm{0.490\pm 0.003}$ & $\bm{0.398\pm 0.004}$ & $\bm{0.346\pm 0.002}$ & $\bm{0.287\pm 0.004}$ & $\bm{0.264\pm 0.004}$ \\
		Ours(BERT) & $0.430\pm 0.008$ & $0.321\pm 0.010$ & $0.294\pm 0.005$ & $0.228\pm 0.007$ & $0.206\pm 0.005$ \\
		Ours(no CNN) & $0.467\pm 0.003$ & $0.357\pm 0.005$ & $0.315\pm 0.002$ & $0.247\pm 0.005$ & $0.227\pm 0.005$ \\
		Ours(no alignment) & $0.478\pm 0.005$ & $0.387\pm 0.004$ & $0.335\pm 0.003$ & $0.271\pm 0.005$ & $0.242\pm 0.004$ \\
		Ours(no augmentation) & $0.381\pm 0.003$ & $0.317\pm 0.006$ & $0.275\pm 0.004$ & $0.229\pm 0.007$ & $0.211\pm 0.003$ \\
		Ours(no homology)  & $0.473\pm 0.003$ & $0.387\pm 0.007$ & $0.341\pm 0.003$ & $0.285\pm 0.010$ & $0.257\pm 0.004$ \\
		Ours(no perturbation) & $0.452\pm 0.002$ & $0.370\pm 0.007$ & $0.320\pm 0.005$ & $0.269\pm 0.005$ & $0.240\pm 0.006$ \\
		\hline
	\end{tabular}
\end{table}

\begin{table}
	\tiny
	\centering
	\caption{(BRI metric) Average IOU of binding residues against the detected ligands. The mean and standard deviation are calculated based on the metric values for 5 different cross-validation folds.}
	\label{t3}
	\begin{tabular}{|c|c|c|c|c|c|}
		\noalign{\smallskip}\noalign{\smallskip}\hline & scPDB(held-out) & COACH420 & HOLO4k & CHEN-holo & CHEN-apo \\
		\hline 
		DeepSurf & $0.402\pm 0.010$ & $0.419\pm 0.013$ & $0.330\pm 0.007$ & $0.372\pm 0.019$ & $0.336\pm 0.020$ \\
		Kalasanty & $0.356\pm 0.000$ & $0.362\pm 0.000$ & $0.344\pm 0.000$ & $0.333\pm 0.000$ & $0.323\pm 0.000$ \\
		Deeppocket & $0.595\pm 0.002$ & $0.506\pm 0.005$ & $0.371\pm 0.004$ & $0.395\pm 0.008$ & $0.382\pm 0.006$ \\
		\hline
		Ours & $\bm{0.643\pm 0.004}$ & $\bm{0.585\pm 0.007}$ & $\bm{0.415\pm 0.002}$ & $\bm{0.495\pm 0.010}$ & $\bm{0.473\pm 0.004}$ \\
		Ours(BERT) & $0.567\pm 0.012$ & $0.459\pm 0.013$ & $0.356\pm 0.007$ & $0.353\pm 0.016$ & $0.328\pm 0.010$ \\
		Ours(no CNN) & $0.624\pm 0.002$ & $0.548\pm 0.004$ & $0.395\pm 0.002$ & $0.450\pm 0.014$ & $0.419\pm 0.006$ \\
		Ours(no alignment) & $0.637\pm 0.005$ & $0.572\pm 0.001$ & $0.413\pm 0.002$ & $0.479\pm 0.005$ & $0.450\pm 0.006$ \\
		Ours(no augmentation) & $0.522\pm 0.001$ & $0.475\pm 0.007$ & $0.347\pm 0.004$ & $0.397\pm 0.010$ & $0.380\pm 0.007$ \\
		Ours(no homology) & $0.628\pm 0.003$ & $0.567\pm 0.009$ & $0.412\pm 0.002$ & $0.481\pm 0.007$ & $0.449\pm 0.010$ \\
		Ours(no perturbation) & $0.596\pm 0.006$ & $0.547\pm 0.008$ & $0.391\pm 0.004$ & $0.468\pm 0.008$ & $0.445\pm 0.008$ \\
		\hline
	\end{tabular}
\end{table}

\begin{figure}%
    \centering
    {{\includegraphics[width=5cm]{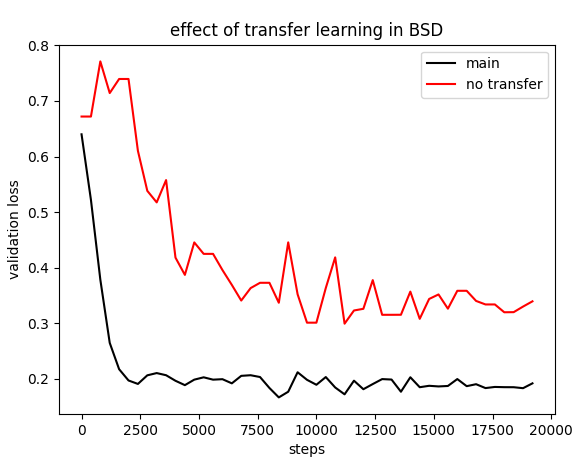} }}
    
    \caption{The effect of transfer learning on the BSD training %
    }
    \label{transfer_curves}%
\end{figure}

\begin{figure}%
    \centering
    {{\includegraphics[width=15cm]{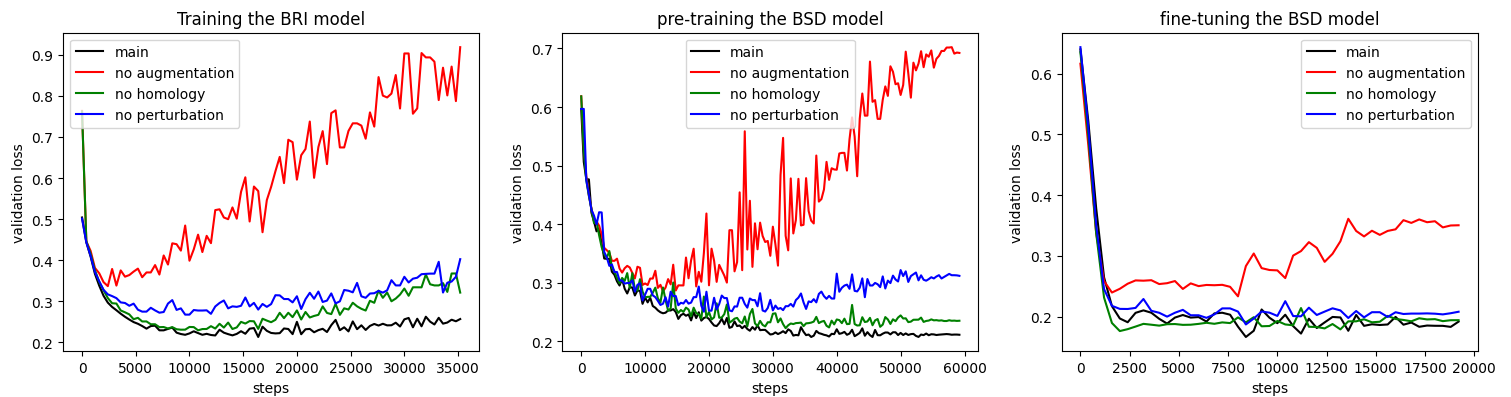} }}
    
    \caption{The effect of the augmentation methods on the training %
    }
    \label{overfitting_curves}%
\end{figure}

The experiment results are summarized in \tableref{t1}, \tableref{t2}, \tableref{t3}, \figref{transfer_curves} and \figref{overfitting_curves}. 

\subsubsection{BSD performance}\label{subsubsection:metric1_result}

In general, \tableref{t1} shows that our method significantly outperformed the baseline methods in the BSD task. Although there is an exception that Deeppocket outperformed our method on CHEN-holo, the gap in performance ($1.2\%p$) is relatively insignificant compared to the average gain in performance ($3.1\%p$) across other datasets. 

%This is true especially when considering the standard deviation. 

The ablation results show that all components had positive effects on the performance in general. In particular, (1) the use of geometric information in the attention layers, (2) the transfer learning and (3) augmentation (in particular the homology augmentation) showed positive contributions consistently and significantly. Omitting each component resulted in the decrease in the model's performance as follows:
\begin{itemize}
\item The geometric information: $4.3\%p$
\item The transfer learning: $4.64\%p$
\item The augmentation as a whole: $3.84\%p$
\item The homology augmentation: $1.74\%p$
\item The random perturbation augmentation: $0.14\%p$ 
\end{itemize}

However, we observed that not all components of our proposed model showed consistent contributions to the model’s performance across different datasets. This was the case for the use of CNN, the grid alignment and the random perturbation. The questionable effects of the first two components may be accounted to the insufficient size of the training dataset for the BSD task. We believe that the model required a larger training dataset to compensate for the increased model complexity from the introduction of the CNN component. Similarly, aligning the grids rather than applying a  random rotation to the grids may have decreased the diversity of the training dataset. Lastly, the random perturbation’s questionable contribution may be attributed to the offset in the effects from the homology augmentation. Indeed, the random perturbation resulted in a significant improvement in performance in the absence of the homology augmentation, which can be confirmed by comparing ``no augmentation" with ``no homology" in \tableref{t1} and \figref{overfitting_curves}. 

\subsubsection{The effect of transfer learning}

\figref{transfer_curves} shows the effect of transfer learning in the BSD training. According to the plot, the transfer learning had two effects on the training process. Firstly, it significantly accelerated the convergence speed. In fact, validation loss almost dropped to the convergence level in 2000 steps. Note that until 4000 steps, we updated only the weights of the un-pretrained parts. Secondly, the transfer learning also significantly improved the validation loss of the converged state. This finding is compatible with the ablation result of \tableref{t1}. 

\subsubsection{BRI performance}

%should revise this

In terms of the BRI performance, \tableref{t2} and \tableref{t3} show that our model outperformed the baselines by a significant margin on all test datasets. In particular, while the strongest baseline model Deeppocket performed poorly on the external test datasets, our model did not. This shows our model generalizes well to the proteins such that no similar proteins are encountered during training. 

The ablation results show that all key components of our method explained in \ref{section:components} contributed significantly in the good performance. These components are (1) the architectural aspects (the CNN and attention layers), (2) the grid alignment process and (3) the augmentation methods.

\subsubsection{The effect of augmentation}

\figref{overfitting_curves} shows that the augmentation contributed significantly in alleviating over-fitting in all training scenarios (pre-training the BSD module, fine-tuning the BSD module and training the BRI module). While the augmentation as a whole was shown to dramatically reduce the over-fitting, individual augmentation methods (random perturbation and homology augmentation) were also shown to be effective. This finding is compatible with the ablation results of \tableref{t1}, \tableref{t2} and \tableref{t3}. 

\subsection{Conclusions}

In general, our model significantly outperformed the baseline methods in terms of both BSD and BRI. All key components in \ref{section:components} contributed to the performance, with some exceptions with respect to BSD. Additional analyses confirmed that the transfer learning and the augmentation worked as intended during training. The transfer learning accelerated and improved the convergence, and the augmentation methods helped overcome the over-fitting problem. 

%Our model's improvement over the baselines were more significant for BRI, as shown in \tableref{t2} and \tableref{t3}. This good BRI performance would make our model useful in \textit{rational drug design}. For example, the identified key interacting residues in a receptor may be used to prioritize docking results (\cite{large_scale_docking}, \cite{orphan_gpcr}). \red{SAR example} 

\section{A case study}
We conducted a case study to demonstrate our model's applicability in real-world drug discovery scenarios. We chose Human Serum Albumin (HSA) as the target protein, due to its relevance to the drug discovery and the fact that it binds to various molecules at different sites (\cite{all_about_albumin}, \cite{hsa_survey}, \cite{hsa_jmpark}). To examine our model's performance on HSA, we drew on two prior studies on the binding sites of HSA (\cite{hsa_survey}, \cite{hsa_oncology_drugs}), and analyzed how compatible our BSD and BRI modules' predictions were with the findings of the studies. 

The remaining parts of this section are organized as follows:
\begin{itemize}
    \item In \ref{case_study:structure_and_BS_of_HSA}, we briefly explain the structure of HSA, and elucidate what the reference papers (\cite{hsa_survey}, \cite{hsa_oncology_drugs}) revealed about the binding sites of it. 
    \item In \ref{case_study:basic_settings}, we explain basic settings of our case study experiments. 
    \item In \ref{subsection:casestudy_bsd}, we analyze our BSD module's performance on HSA based on the findings of \cite{hsa_survey}.
    \item In \ref{subsection:casestudy_bri}, we analyze our BRI module's performance on HSA based on the findings of \cite{hsa_oncology_drugs}.
\end{itemize}

\subsection{The structure and Binding Sites of HSA}\label{case_study:structure_and_BS_of_HSA}

\begin{figure}%
    \centering
    \includegraphics[width=8cm]{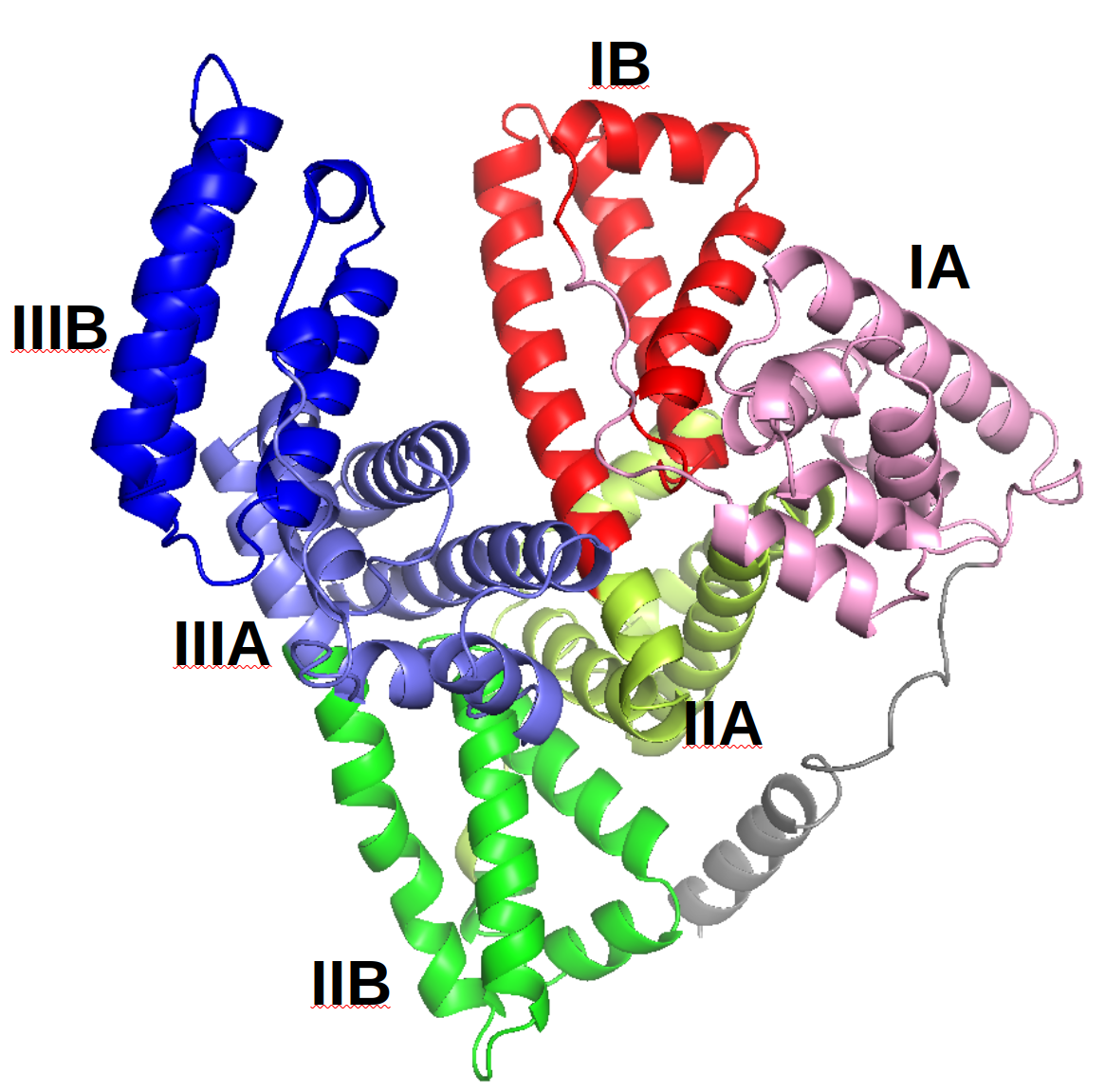} %
    \caption{
    The subdomains of HSA
    }
    \label{hsa_subdomains}%
\end{figure}

HSA is composed of three homologous domains (I, II, III), each composed of two subdomains A and B, as depicted in \figref{hsa_subdomains} (\cite{all_about_albumin}, \cite{hsa_survey}). The authors of \cite{hsa_survey} performed a large-scale survey of HSA binding sites based on 142 crystal structures involving HSA. By analyzing the complexes, they identified 14 different binding sites, alongside their \textit{frequencies}. They noted that binding sites ``IB", ``IIA" and ``IIIA" dominated clearly. Inspired from this result, the authors of \cite{hsa_oncology_drugs} provided more detailed analysis on the binding sites at the IB subdomain. They did so by inspecting the crystal sructures of 6 oncology drugs (9-amino-camptothecin, camptothecin, idarubicin, teniposide, etoposide and bicalutamide) in complex with HSA. In particular, for each structure, they identified all key residues of HSA and their interaction types with the drug molecule. 

\subsection{Basic settings}\label{case_study:basic_settings}
We designed the analyses such that they can faithfully
assess our model's real-world applicability. 

Firstly, before the analyses, we trained our model with a new dataset split (different from the ones used in our main experiments) to prevent data leakage. Specifically, we removed all 42 ``albumin" structures from the scPDB v.2017 dataset, randomly sampled a validation set of size 1000 from the remaining, and took all the other proteins as the training set. Moreover, we ensured that there is no leakage coming from the homology-based augmentation, by using a new augmentation dataset generated from the new training set.

Secondly, all our model's predictions are based on taking as input the HSA structure provided by the Alphafold database (\cite{alphafold_human_proteome}. Therefore, a good performance under this setting would imply that our model can be used to predict the binding sites of a protein without any known experimental structures. In particular, one can make use of the publicly available Alphafold database in the predictions. 

\subsection{Binding Site Detection}\label{subsection:casestudy_bsd}

\begin{figure}%
    \centering
    \includegraphics[width=12cm]{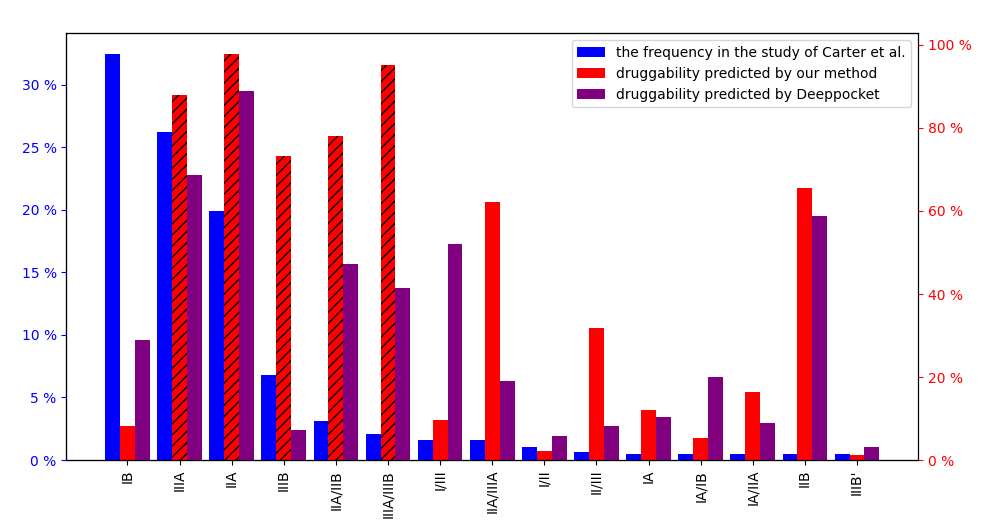} %
    \caption{
    A comparison between our BSD module's predictions and the \textit{frequencies} from \cite{hsa_survey}
    }
    \label{case_study:rank_plot}%
\end{figure}

\subsubsection{experiment procedure} 

First, we obtained our model and Deeppocket's BSD module's predictions on 15 binding sites identified by \cite{hsa_survey}. This means that the predictions were made based on 15 different inputs, where we set the ``binding site center'' to be the mean alpha carbon coordinates of residues comprising one of the binding sites. The indices of residues comprising each binding site were provided by \cite{hsa_survey}. Note that we re-trained the Deeppocket model with a new dataset split to avoid data leakage, just as we did for our model. 

Then, we assessed each model's predictions by comparing them with the ranks of the \textit{frequencies} of the binding sites as recorded in \cite{hsa_survey}. 

\subsubsection{results and analysis}
The results are summarized in \figref{case_study:rank_plot}. Note that our model successfully assigned high (more than $70\%$) druggability scores on the second to sixth most \textit{frequent} binding sites. Moreover, these five were exactly the binding sites that scored the highest. The probability of the latter condition being met by a random ordering is only $0.2\%$, which shows the statistical significance of our model's ability in replicating binding sites' \textit{frequency} ranks. On the other hand, although Deeppocket assigned high druggability scores on the second and third most \textit{frequent} binding sites, it failed to do so on the fourth to sixth most \textit{frequent} ones.

\subsection{Binding Residue Identification}\label{subsection:casestudy_bri}

\begin{figure}%
    \qquad
    \subfloat[\centering All 6 drug molecules examined in \cite{hsa_oncology_drugs}]{{\includegraphics[width=4cm]{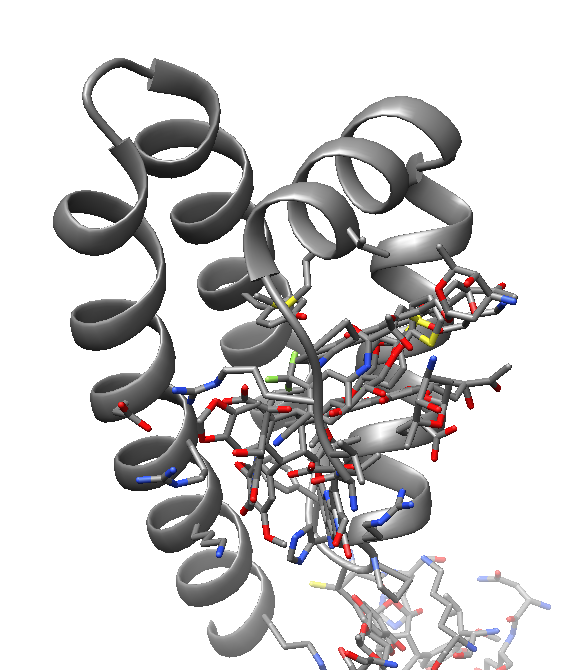} }}%
    \qquad
    \subfloat[\centering Our model's prediction]{{\includegraphics[width=4cm]{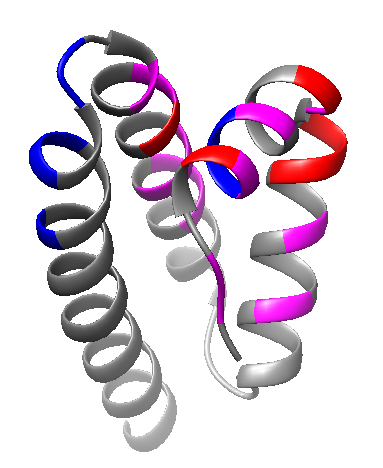} }}%
    \qquad
    \subfloat[\centering Deeppocket's prediction]{{\includegraphics[width=4cm]{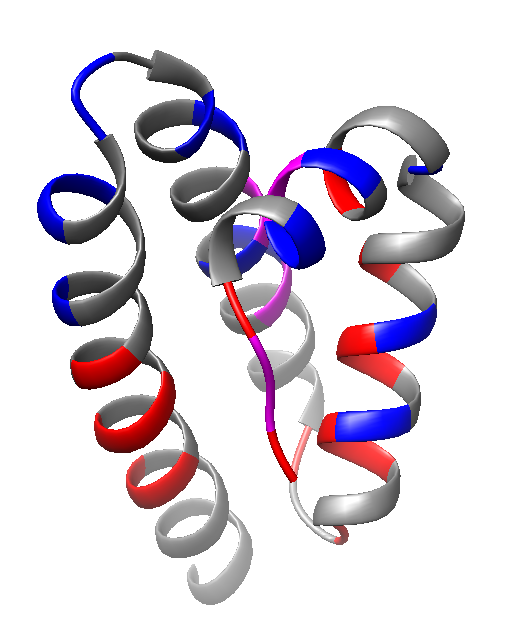} }}%
    \qquad
    \subfloat[\centering Kalasanty's prediction]{{\includegraphics[width=4cm]{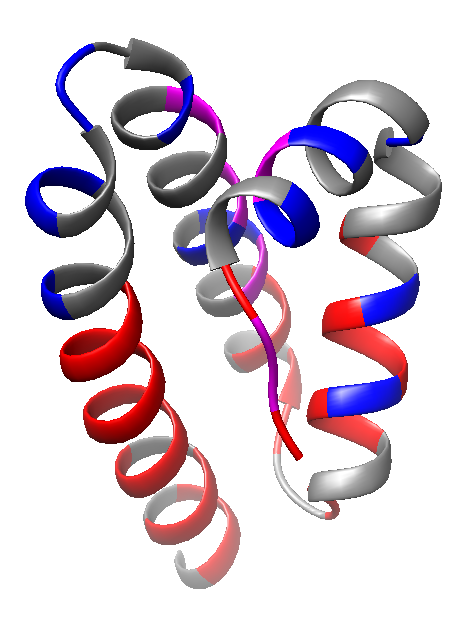} }}%
    \qquad
    \subfloat[\centering DeepSurf's prediction]{{\includegraphics[width=4cm]{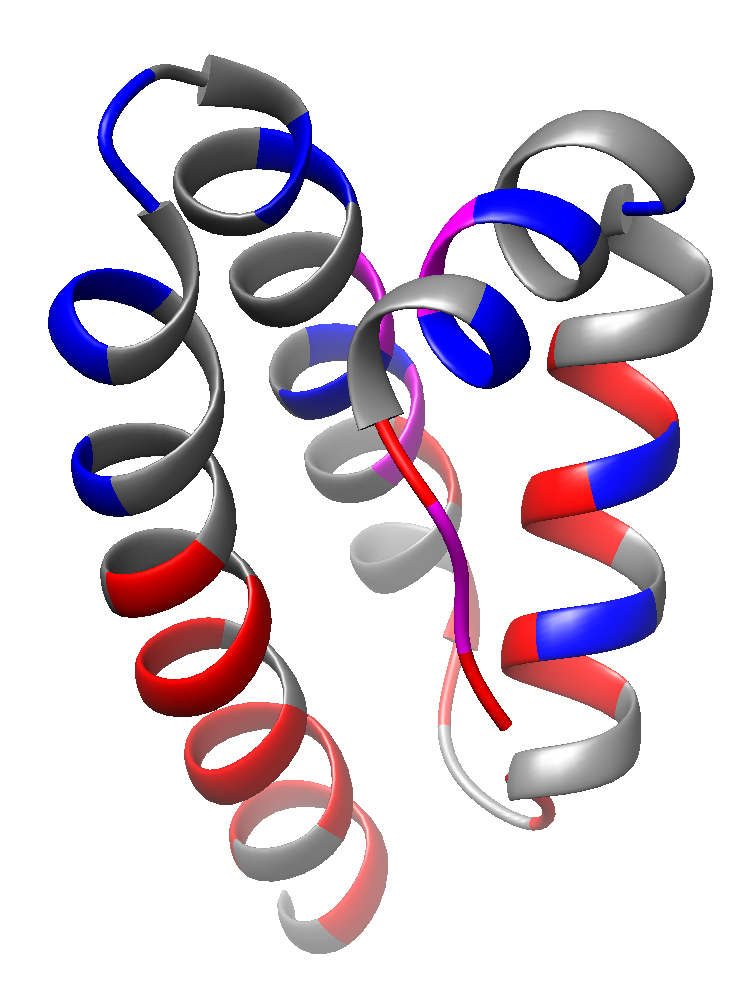} }}%
    \caption{
    The purple, red and blue residues indicate the true-positive, false-positive and false-negative binding site residues respectively. Therefore, the larger the purple region is compared to the regions with the other colors, the better the prediction is. 
    }
    \label{case_study:bsite_visuals}%
\end{figure}

\subsubsection{experiment procedure}
First, we predicted the binding site residues in the IB subdomain of HSA using our model and all the baseline methods --- Deeppocket, Kalsanty and DeepSurf. When obtaining our model's (resp. Deeppocket's) predictions, we set the ``binding site center" to be the mean alpha carbon coordinate of the IB subdomain residues, and run the BRI module (resp. the \textit{segmentation model}). Note that as we did for our model, we re-trained the Deeppocket and DeepSurf models with a new dataset split to avoid data leakeage.

Then, we visually inspected the binding residues predicted by different methods, comparing them with the ground truth provided by \cite{hsa_oncology_drugs}. As the ground truth, we took the union of all the ``key residues" (in the IB subdomain) of the 6 drug molecules as determined by \cite{hsa_oncology_drugs}. 

\subsubsection{results and analysis}
The results are visualized in \figref{case_study:bsite_visuals}. From the figures, it is clear that our model's prediction best matched with the ground truth. In particular, while all the baseline models resulted in false-positives within the left-bottom helix, our model did not. Also, while all the baseline models showed limited precision in identifying the key residues in the other parts of the IB subdomain, our model showed mostly successful identifications.

\section{Methods}\label{section:methods}
\subsection{The residue orientation}\label{residue_orientation}

Our model uses a concept of ``residue orientation" in the grid alignment process and the geometric attention layers. In the grid alignment process, the grid axes are aligned with respect to the orientations. In the geometric attention layers, the orientations are used as a part of the input.

We define the residue orientation in terms of the relative positions of atoms surrounding the alpha carbon, as in \cite{alphafold}. More precisely, when $\textbf{x}_1$, $\textbf{x}_2$ and $\textbf{x}_3$ are the coordinates of $N$, $C\alpha$ and $C$ (the carbon that is not $C\beta$ and adjacent to $C\alpha$), the rotation matrix $R=(\textbf{e}_1\enspace \textbf{e}_2\enspace \textbf{e}_3)$ that we refer to as ``orientation" is obtained as follows:
\begin{align*}
	&\textbf{v}_1 = \textbf{x}_3 - \textbf{x}_2\\
	&\textbf{v}_2 = \textbf{x}_1 - \textbf{x}_2\\
	&\textbf{e}_1 = \textbf{v}_1 / \norm{\textbf{v}_1}\\
	&\textbf{u}_2 = \textbf{v}_2 - (\textbf{e}_1\cdot\textbf{v}_2)\textbf{e}_1\\
	&\textbf{e}_2 = \textbf{u}_2 / \norm{\textbf{u}_2}\\
	&\textbf{e}_3 = \textbf{e}_1 \times \textbf{e}_2 \\
\end{align*}

\subsection{The grid featurization}\label{subsection:methods/grid_featurization}

As explained in \ref{subsection:bsd_module} and \ref{subsection:bri_module}, our BSD and BRI modules take as input a sequence of 3D voxelized images, where each image represents the local environment of a residue. The process of making such voxelized images, the ``grid featurization", proceeds as follows:

\begin{enumerate}
\item Collect coordinates $\textbf{x}_i\in\mathbb{R}^3$ ($1\le i\le n$) of the heavy (non-hydrogen) atoms in the protein. 
\item Collect feature vectors $f_i\in\mathbb{R}^{d_{feature}}$ ($1\le i\le n$) of the heavy atoms.
\item Given a choice of \textit{grid axes} $(\textbf{e}_1,\textbf{e}_2,\textbf{e}_3)$, a cubical grid is laid such that the centers of the voxels in the grid become 
\begin{equation*}
    t + \sum_{i=1}^{3}r(a_i-\frac{L-1}{2})\textbf{e}_i\quad \text{($(a_1, a_2, a_3)\in\set{0,\cdots,L-1}^3$)}
\end{equation*}
, where $t$ is the grid center (the alpha carbon coordinate), $r$ is the \textit{grid resolution} and $L$ is the \textit{grid size}.  
\item Compute feature vectors corresponding to each voxel in the grid, by summing up those of nearby heavy atoms. 
\end{enumerate} 

The same process was used in \cite{Deeppocket}, \cite{DeepSurf} and \cite{Kalasanty}, although the grids in those methods are not laid on the protein residues. 

The specific details of our grid-featurization process are:
\begin{itemize}
\item The grid resolution is $1\AA$ and the grid size is $16$. 
\item The atom features are of dimension 18 and include atom types, hybridization, degree, partial charge and aromaticity. (\cite{Kalasanty}, \cite{Deeppocket}.)
\end{itemize}

\subsection{The grid alignment}

As explained in \ref{subsection:se3_invariance}, we apply the ``grid alignment" during the grid featurization to promote SE(3)-invariance. This means that we choose the grid axes to be the vectors comprising the residue orientation (See \ref{residue_orientation}). 

%Moreover, when we apply the random perturbation augmentation (\ref{augmentation_strategies}), we alter the grid axes by applying a perturbing rotation. When $P$ is the perturbation matrix (the matrix corresponding to the perturbing rotation), we take the grid axes to be $(\textbf{e}_1',\textbf{e}_2',\textbf{e}_3')$, where 
%\begin{equation*}
%    \begin{pmatrix}\textbf{e}_1'&\textbf{e}_2'&\textbf{e}_3'\end{pmatrix}
%    =P\begin{pmatrix}\textbf{e}_1&\textbf{e}_2&\textbf{e}_3\end{pmatrix}
%\end{equation*}

\subsection{The CNN model}
Our model incorporates a 3D CNN to encode the residue-level grid features. For the CNN architecture, it uses a 3D BottleNeck ResNet model introduced in \cite{DeepSurf}. The model is adapted from the bottleneck ResNet model introduced in \cite{resnet} for the image classification. The bottleneck architecture reduces the number of parameters thus enables the use of a deeper network. \cite{DeepSurf} showed that the 3D BottleNeck ResNet model, while being light, performed competitively well compared to its non-bottleneck counterpart. 
%We used the same hyper-parameters as \cite{DeepSurf}. 

\subsection{The geometric attention layers}\label{rayatt_method}

The geometric attention is an integral part of our model's architecture. The attention mechanism was adopted from Alphafold(\cite{alphafold})'s IPA(Invariant Point Attention) with an adjustment necessary to adapt it to our forms of inputs.

The inputs of the attention layers are composed of the following:
\begin{itemize}
\item $x_i\in\mathbb{R}^{d_{hidden}}$ ($i=1,\cdots,n$), hidden vectors associated to the residues.
\item $T_i=(R_i, t_i)\in SO(3)\times\mathbb{R}^3$  ($i=1,\cdots,n$), the \textit{local frames} associated to the residues, where $t_i$ is the position of the alpha carbon and $R_i$ is the rotation matrix that represents the residue orientation (See \ref{residue_orientation}). Note that the operation $v\mapsto T_i v$ maps the local coordinates (with respect to the local frame) to the corresponding global coordinates, and the operation $u\mapsto T_i^{-1} u$ plays the reverse role. 
\end{itemize}

Then, the computation is carried out in the following steps (See also \figref{att_figure}):
\begin{enumerate}
	\item The standard query and key vectors $q_i^h$ and $k_i^h$ are computed by the linear mappings from $x_i$. Here, $h$ stands for a ``head". 
	\item The geometric query and key vectors $\textbf{q}_i^{hp}$ and $\textbf{k}_i^{hp}$ in $\mathbb{R}^3$ are computed by the linear mappings from $x_i$. Here, $h$ stands for a ``head" and $p$ stands for a ``point" of attention. 
	\item The \textit{attention weight} from the $i$-th token to the $j$-th token is computed from a linear combination of the standard attention weight 
	\begin{equation*}
	w_{ij}^{h,standard }=\frac{1}{\sqrt{d_{hidden}}}q_i^h\cdot k_j^h
	\end{equation*}
	 and the geometric attention weight 
	 \begin{equation}\label{geometric_attention_term}
	 w_{ij}^{h,geometric}=\frac{1}{\sqrt{N_{points}}}\sum_{p}\norm{T_i\textbf{q}_i^{hp} -T_j\textbf{k}_j^{hp}}
	 \end{equation} by applying a softmax operation. More precisely, the attention weight becomes 
 	\begin{equation*}
 		w^h_{ij} = softmax_{j} (\frac{1}{\sqrt{2}}(w_{ij}^{h,standard}-\log(1+\gamma^h)w_{ij}^{h,geometric}))
 	\end{equation*}
 	where $\gamma^h$ is a learnable parameter. 
 	\item The standard value vectors $v_j^h$ are computed by a linear map from $x_j$, and aggregated as 
 	\begin{equation*}
 		o_i^h=\sum_{j}w_{ij}^hv_j^h 
 	\end{equation*} 
 	\item The geometric value vectors $\textbf{v}_j^h$ are computed by a linear map from $x_j$, and aggregated as 
 	\begin{equation}\label{geometric_value_aggregation}
 		\textbf{o}_i^{hp} = T_i^{-1} ( \sum_{j}w_{ij}^hT_j\textbf{v}_j^{hp})
 	\end{equation} 
 	\item The aggregated vectors as well as their sizes are concatenated and linearly mapped via $f_{final}$ to produce the output of the attention layer 
 	\begin{equation*}
 		x_i'= f_{final}(concat_{h,p}(o_i^h,\textbf{o}_i^{hp},\norm{\textbf{o}_i^{hp}}))
 	\end{equation*}
\end{enumerate}

The adjustment made to the original IPA is the omission of the ``attention bias" term. In the original paper, this term was based on the ``pair representation" computed at the earlier stages of the Alphafold architecture using the evolutionary information of the protein. Since our model does not involve this representation, the omission is necessary to use the IPA in our model.

%After the attention layer, there comes a pointwise feed-forward layer. Both of these are followed by dropout,  residual connection and layer normalization.  

\subsection{SE(3)-invariance of the geometric attention layers}

As noted in \ref{subsection:se3_invariance}, our model's SE(3)-invariance relies on the attention layer's SE(3)-invariance. 
In order to prove that the layer is SE(3)-invariant, one has to examine the equality \eqref{att_se3_invariance}. Essentially, the equality holds for our attention layer because the quantities \eqref{geometric_attention_term} and  \eqref{geometric_value_aggregation}, as functions of $\set{x_i}_{i=1}^n$ and $\set{T_i}_{i=1}^n$, satisfy the same equality. For any $T\in SE(3)$, the \eqref{geometric_attention_term} satisfies the equality because $\norm{TT_i\textbf{q}_i^{hp} -TT_j\textbf{k}_j^{hp}}=\norm{T(T_i\textbf{q}_i^{hp} -T_j\textbf{k}_j^{hp})}=\norm{T_i\textbf{q}_i^{hp} -T_j\textbf{k}_j^{hp}}$ (since $T$ preserves the norm), and \eqref{geometric_value_aggregation} satisfies the equality because $(TT_i)^{-1} ( \sum_{j}w_{ij}^h(TT_j)\textbf{v}_j^{hp})=T_i^{-1} T^{-1}T(\sum_{j}w_{ij}^hT_j\textbf{v}_j^{hp})=T_i^{-1} ( \sum_{j}w_{ij}^hT_j\textbf{v}_j^{hp})$. Note that this derivation is almost identical to that presented in \cite{alphafold}(Supplementary Information, page 28).

%We used the same default hyper-parameters as Alphafold:
%\begin{itemize}
%	\item hidden dimension: 128
%	\item intermediate dimension of the pointwise feed-forward network: 512
%	\item number of attention heads: 12
%	\item dimension of query and key vectors: 16,
%	\item dimension of value vectors: 16
%	\item number of attention points: 4
%	\item dropout rate: 0.1
%\end{itemize}
%For the parts shared by the BRI and BSD module architectures, we used 3 attention layers. For the BSD module, we used one additional layer. 

%\subsection{Reference Point Embedding}
%For the BSD module, we additionally used ``reference point embedding", which informs each residue of the location of the center of the binding site candidate. More precisely, when $\textbf{v}$ is the center, 
%$f(R_i^{-1}(\textbf{v}-\textbf{t}_i))$ is added to the hidden vectors in the BSD module right after the parts shared with the BRI module, where $f$ is a learnable linear function. \blue{This enables differentiating the hidden vectors based on the residues' relative location to the center. This is analogous to the use of ``position embedding" of BERT (\cite{vaswani}), which differentiates the hidden vectors based on the token's index. The location relative to the center is relevant in our case, because our BSD module's optimization goal is  strongly tied with the location of the center (\ref{subsection:evaluation}). Moreover, the embedding takes this specific form to retain the SE(3)-invariance of the model.} 

\subsection{The Homology-based Augmentation}

\begin{figure}
    \centering
    \includegraphics[width=18cm]{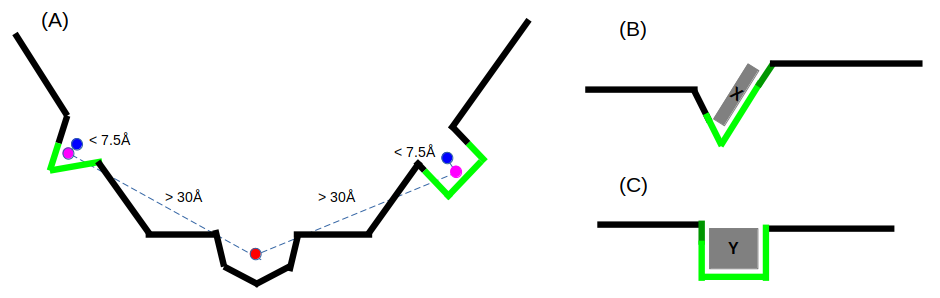}
    \caption{
    These figures illustrate the way our homology-based augmentation determines the positive and negative binding site candidates in augmented proteins. 
    Figure (A) depicts an augmented protein, where there are two positive (the blue points) and one negative (the red point) binding site candidate centers. Out of binding site candidates proposed by Fpocket, they are labeled based on the distances to the \textit{proxy centers} (the purple points) of binding sites inferred from homology relations. Figures (B) and (C) depict the homologous proteins in the original database that gave rise to the inferred binding sites in the augmented protein. $X$ and $Y$ are their ligands. The bright and dark green regions of the chains indicate the residues in close proximity to the ligands, while only the bright green region has evolutionary correspondence to residues in the augmented protien. The bright green region must comprise at least 50\% of the entire green region in order for the binding site to count. 
    }
    \label{aug_visual1}
\end{figure}

\begin{figure}%
    \qquad
    \subfloat[\centering structure of UniProt Q9VC32 from the Alphafold database]{{\includegraphics[width=5cm]{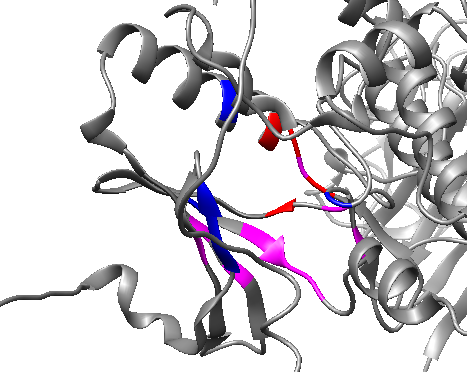} }}%
    \centering
    \subfloat[\centering structure of PDB 4G34 from scPDB]{{\includegraphics[width=5cm]{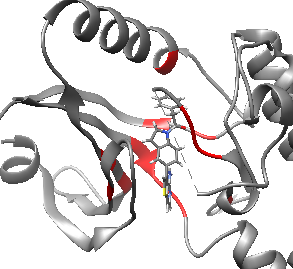} }}%
    \qquad
    \subfloat[\centering structure of PDB 4BID from scPDB]{{\includegraphics[width=5cm]{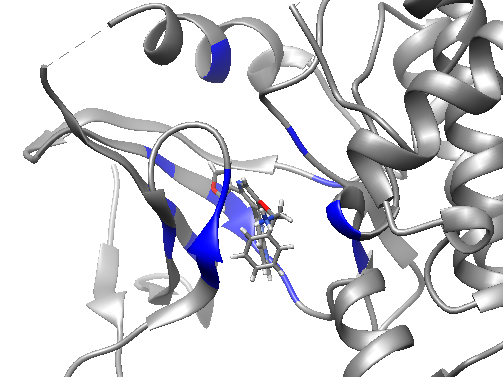} }}%
    \caption{
    These figures illustrate the way our homology-based augmentation assigns residue labels in augmented proteins with respect to a positive binding site candidate.
    Figure (A) illustrates UniProt protein Q9VC32 as an augmented protein. The red and purple residues correspond to the red residues in figure (B) of the PDB protein 4G34, which are the ligand-binding residues. Similarly, the blue and purple residues correspond to the blue residues in figure (C) of the PDB protein 4BID, which are the ligand-binding residues.  
    The purple residues, the intersection, attain labels 1.0, while the other colored residues attain labels 0.5. This means that our augmentation method regards the purple residues as the most likely ligand-binding ones.%
    }
    \label{aug_visuals}%
\end{figure}

% In general it is hard to read this section. The flow is a bit messy.

The ``homology-based augmentation" is one of our new augmentation methods used to overcome the problem of over-fitting. It is distinguished from the usual augmentation methods in that it is not based on transformations applied to the samples on the fly during the training. Instead, it pre-computes appropriate ``augmented samples" out of an external database of unlabelled protein structures, and use the augmented dataset consisting of the augmented samples during training. Essentially, the augmented samples are selected based on the sequence alignments computed with respect to the proteins in the original training
set. In this section, we describe the augmentation method in more detail, clarifying its inputs, outputs, the procedures and the underlying rationale.

The augmentation method requires a \textit{seed} database $\mathcal{S}^*$ of multi-chain protein-ligand complexes and a \textit{target} database $\mathcal{T}$ of single-chain protein structures. In our instantiation, $\mathcal{S}^*$ was the portion of the scPDB dataset used to train the current fold of the cross-validation. 
For $\mathcal{T}$, we used the entire Alphafold Protein Structure Database (as of April. 2022) that contained 992,316 protein structures from the proteome of human and 47 other key organisms as well as the Swiss-Prot entries.

The augmentation procedure outputs
two types of information, which together form the ``augmented dataset" and are used during the training as described in \ref{applying_augmentations}. The first information denotes the centers of the binding site candidates in proteins in a selected subset of $\mathcal{T}$, labeled either \textit{positive} or \textit{negative}. This is used to augment the BSD train dataset. The second information denotes, for each previous positive binding site candidates, the likelihood of each nearby protein residue to be a ligand-binding residue. This is used to augment the BRI train dataset.

In the following, we describe the steps of the procedure. The \textit{italized words} are general terms whose specification may vary depending on own's needs. 
%We provide our specification in the following paragraph.  
Whenever there is an \textit{italized word}, we provide our specification at the end of the step. 
\begin{itemize}
\item 
[(1)] In each holo structure of $\mathcal{S}^*$, find ligands \textit{associated to} exactly one chain. As a result, obtain a database $\mathcal{S}$ of protein chains associated to at least one such single-chain ligands. (A chain can be associated to multiple single-chain ligands) We define that a chain and a ligand are \textit{associated to} each other if they have heavy atoms within $4\AA$ to each other. 

\item
[(2)] Run a \textit{homology search algorithm} with $\mathcal{S}$ as the query database and $\mathcal{T}$ (the database of single-chain protein structures) as the target database. Based on the results, obtain a MSA for each chain in $\mathcal{S}$. For the \textit{homology search algorithm}, we use the software HHBlits with its default setting. 

\item [(3)] 
For each triplet $(x, l, y)$, composed of: 
\begin{enumerate}
\item a query chain $x$ in $\mathcal{S}$ 
\item a ligand $l$ associated to $x$ found in step 1 of the procedure and 
\item a target chain $y$ aligned with $x$ in the MSA, 
\end{enumerate}
determine whether the ligand $l$’s binding site in $x$ is \textit{preserved} in $y$. The triplets for which the previous determination was affirmative will be called \textit{preserving}. We define that a triplet $(x,l,y)$ is \textit{preserving} if at least half of the residues of $x$ that are in close contact with $l$ (heavy atoms within $4\AA$) are aligned with a residue of $y$ in the MSA.

\item [(4)]
For each preserving triplet $(x,l,y)$, find a \textit{proxy center} of the binding site in $y$ that corresponds to the ligand $l$'s binding site in $x$. We define the \textit{proxy center} to be the mean of the alpha carbon coordinates of the residues of $y$ aligned in the MSA with a residue of $x$ in close contact with $l$.
\item [(5-1)]
On each chain $y$ in $\mathcal{T}$ that is involved in at least one preserving triplet, run Fpocket to get an initial list of binding site center candidates. Label a candidate center ``positive" if it is within a \textit{lower threshold} from a proxy center obtained in the previous step. Label it ``negative" if it is farther than a \textit{higher threshold} from any such proxy center. If a candidate center does not fall into these categories, ignore it and exclude it from the dataset. 
We define the \textit{lower threshold} to be $7.5\AA$ and the \textit{upper threshold} to be $30\AA$.  \figref{aug_visual1} illustrates this step using schematic figures.

\item [(5-2)]
For each positively labeled binding site candidate from the previous step, label residues of $y$ with the estimated likelihood of comprising the binding site. The estimate is obtained as a result of ``voting" of the homologous chains in $\mathcal{S}$ that gave rise to the binding site. More specifically, among the preserving triplets $(x,l,y)$ whose proxy center gave rise to the binding site (in the sense of the step (5-1)), the proportion of such triplets for which the residue at hand corresponds (in MSA) to a residue in the binding site of $l$ is computed. \figref{aug_visuals} illustrates this step using an actual example. 
\end{itemize}

The assignments of different labels in the previous procedure are based on the following hypotheses:
\begin{itemize}
\item The positive binding site label: if a pocket-like site (discovered by a geometry-based BSP method) is surrounded by the sequence fragments that are homologous to the binding site sequence fragments of other proteins, it is likely to be a binding site. 
\item The negative binding site label: even if a site is pocket-like, if it is far from any sequence fragments that are homologous to the binding site sequence fragments of other proteins (in a given seed database), it is unlikely to be a binding site.
\item The residue labels: whether a residue nearby a binding site is a part of the binding site or not can be determined by whether the same is true for corresponding residues of the homologous binding sites. 
\end{itemize}

These hypotheses except the second one have been the bases of the \textit{template-based} BSP methods introduced in \ref{subsubsection:template-based_methods}. 

However, the second hypothesis is, to our knowledge, has not been employed in the previous BSP methods. The result of ablating the homology augmentation in \tableref{t1} provides a partial evidence that this hypothesis is valid at least to some degree. Otherwise if the negative labels did not provide valuable learning signals, the homology augmentation would have had only adversary effects on the BSD module's performance.

\bibliographystyle{alpha}
\bibliography{main}

%------------------------------------------------------
\end{document}

\title{Supplementary Information}

\maketitle

\section{Model hyper-parameters}

\subsection{CNN}
We used the same hyper-parameters as \cite{DeepSurf} for the 3D BottleNeck ResNet model. 

\subsection{Attention}

For individual attention layers, we used the same default hyper-parameters as Alphafold:
\begin{itemize}
	\item hidden dimension: 128
	\item number of attention heads: 12
	\item dimension of query and key vectors: 16,
	\item dimension of value vectors: 16
	\item number of attention points: 4
	\item dropout rate: 0.1
\end{itemize}
At the end of each layer, we used a 2-layer point-wise feed-forward network with the intermediate hidden dimension 512. 

For the part of the architecture that BSD and BRI modules share, we used 3 attention layers. For the BSD module, we used one additional attention layer. 

\section{Details of training}

We will summarize the details such as hyper-parameters for each training stage, 

\subsection{Pre-training BSD module}\label{subsubsection:BSD_pre-training_details}

\begin{itemize}
\item Minibatch size: 16 
\item Batch sampling: balanced BSD labels (See \ref{m-subsection:BSD_training})
\item Loss function: balanced binary cross entropy (See \ref{m-subsection:BSD_training}) 
\item Reduction of loss terms from individual datapoints: mean 
\item Peak learning rate: 1e-4 
\item Learning rate schedule: one-cycle cosine scheduling (\cite{cosine_annealing}) with 6000 warmup steps and 60000 total steps 
\item Optimizer: AdamW with $\beta_1=0.9$, $\beta_2=0.999$ and weight decay rate 0.01
\end{itemize}

%We used minibatch size 16, and peak learning rate 1e-4, with the loss terms from individual datapoints mean-reducted. We used AdamW optimizer with $\beta_1=0.9$, $\beta_2=0.999$ and weight decay of rate 0.01 and the cosine learning rate scheduler (the most simple form of cosine annealing \cite{cosine_annealing}) with 6000 warmup steps and 60000 total steps. 

\subsection{Fine-tuning BSD module}\label{subsubsection:BSD_fine-tuning_details}

\begin{itemize}
\item Minibatch size: 16 
\item Batch sampling: balanced BSD labels (See \ref{m-subsection:BSD_training})
\item Loss function: binary cross entropy  
\item Reduction of loss terms from individual datapoints: mean 
\item Peak learning rate: 1e-6 
\item Learning rate schedule: one-cycle cosine scheduling (\cite{cosine_annealing}) with 4000 warmup steps and 20000 total steps 
\item Weight freezing: Up to 8000 steps, the parameters from the part of the architecture shared with the BRI module are frozen (See \ref{m-subsection:BSD_training})
\item Optimizer: AdamW with $\beta_1=0.9$, $\beta_2=0.999$ and weight decay rate 0.01
\end{itemize}

%For a batch, the resulting weighted sums are mean-reduced to obtain the loss.  
%We use minibatch size 16, and peak learning rate 1e-4. We use AdamW optimizer with $\beta_1=0.9$, $\beta_2=0.999$ and weight decay of rate 0.01 and the cosine learning rate scheduler (the most simple form of cosine annealing \cite{cosine_annealing}) with 6000 warmup steps and 60000 total steps.   

%In the second stage, we use mean-reduced binary cross entropy loss, minibatch size 16, peak learning rate 1e-6, AdamW optimizer with $\beta_1=0.9$, $\beta_2=0.999$ and weight decay of rate 0.01, and the cosine learning rate scheduler with 4000 warmup steps and 20000 total steps. We freeze the model parameters of the parts shared by the BRI module architecture until 8000 steps, to promote smoother fine-tuning of the parameters.   

\subsection{Training BRI module}\label{subsubsection:BRI_training_details}

We used the same details as in \ref{m-subsubsection:BSD_pre-training_details}, except the balanced batch sampling. When training BRI module, as explained in \ref{m-subsection:BRI_training}, we used only the binding sites with positive BSD labels. 

\section{Datasets}

In our experiments, we used the datasets scPDB, COACH420, HOLO4K and CHEN. As explained in \ref{m-subsection:datasets},  We used scPDB for a held-out cross validation, and used the other ones as external test datasets.

\subsection{Characteristics of the datasets}
ScPDB (\cite{scPDB}) is a large database of binding sites from the Protein Data Bank. COACH420 (\cite{tmsite_ssite}) consists of 420 single-chain structures from the COACH dataset. HOLO4K (\cite{holo4k_origin}) consists of 4009 larger multi-chain structures. The CHEN dataset (\cite{chen_dataset}) is smaller, but covers a wide range of non-homologous structures. Also, it has 104 holo structures with apo counterparts, where corresponding holo and apo structures have the same SCOP family, 80\% sequence similarity, and TM score above 0.5.  These holo and apo structures resulted in ``CHEN-holo" and ``CHEN-apo" datasets used in our experiments.  

\subsection{Ligands used for evaluations} 

Evaluation of binding sites depends on what ``ligands" are considered to constitute binding sites. ``Biologically irrelevant" ligands such as additives are usually excluded in this calculation. Therefore, we will clarify what ligands we considered for each dataset. For scPDB and CHEN, we used the ones the original databases provided. For COACH420 and HOLO4K, we used the ligands in the PDB files provided by the p2rank (\cite{p2rank}) repository (\cite{p2rank-dataset_repo}), which are those with PDB ligand code listed in ``biologically relevant ligands" curated by MOAD 2013 database (\cite{moad2013_paper}).

\subsection{Structural alignments of apo structures and holo structures}

To evaluate BSP algorithm's performance on the CHEN-apo dataset, we projected the ligands in CHEN-holo to the corresponding protiens of CHEN-apo. We will describe the projection procedure. 

Let $H$ be a protein structure from CHEN-holo, $L_1,\cdots,L_k$ be ligands within $H$, and $P$ be the structure of the corresponding protein in CHEN-apo. We positioned the ligands $L_i$ onto $P$ as follows.  
\begin{itemize}
\item Find a sequence alignment of $H$ and $P$, using the software \textsc{mmseqs}. 
\item  Find a SE(3) transformation $\phi_i$ from the space of $H$ onto the space of $P$ that minimizes the mean square alpha carbon distance between the pairs of residues $(R, R')$ such that
\begin{itemize}
    \item $R$ is from $H$ and $R'$ is from $P$
	\item $R$ and $R'$ are aligned 
	\item $R$ or one of its (two) neighboring residues is within $4\AA$ from $L_i$.  
\end{itemize} 
(If there was no such pair, the ligand was ignored, and not projected to the apo structure)
\item Consider $L_i'=\phi_i(L_i)$ as a ligand within the structure $P$ 
\end{itemize}

\subsection{dataset statistics} 
The procedures described in the previous two sections resulted in the following dataset statistics: 

\begin{table}[h]
	\tiny
	\centering
	\label{dataset_stats}
	\begin{tabular}{|c|c|c|c|c|}
	    \hline 
		dataset & number of proteins & number of binding sites & average number of binding sites & maximum number of binding sites \\
		\hline 
		scPDB & 16612 & 17594 & 1.06 & 4\\ 
		\hline
	    COACH420 & 299 & 494 & 1.65 & 12 \\ 
	    \hline
	    HOLO4K & 3378 & 7141 & 2.11 & 24 \\ 
	    \hline
	    CHEN-holo & 104 & 244 & 2.35 & 11 \\ 
	    \hline
	    CHEN-apo & 104 & 243 & 2.34 & 11 \\
		\hline
	\end{tabular}
\end{table}

\section{Comparison of our evaluation metrics with those of baseline methods}\label{supplementary:metric_pecularities}

As explained in \ref{m-subsubsection:evaluation_metrics}, we used three evaluation metrics for BSP --- \textit{success rate}, \textit{IOU} and \textit{conditional IOU}. Each of these is inspired by previous works, but not exactly same as theirs. 

\subsection{success rate}
This metric is similar to those used in the baseline methods (\cite{Deeppocket}, \cite{Kalasanty} and \cite{DeepSurf}). Compared to their definitions, our definition makes two specific choices. First, as in \cite{Kalasanty}, our definition uses F1 score to compare predicted binding sites and actual ligands in each protein. This differs from the \textit{precision} version used in \cite{Deeppocket} and \cite{DeepSurf}.  These two may differ when more than one \textit{ligand of interest} is bound to the protein. In that case, the \textit{precision} version may give the \textit{perfect} score when all predicted binding sites are nearby a single ligand. %Therefore, we believe that using F1 score is more reasonable for fair evaluations. 
Secondly, in our definition, the definition of ``detection" is based on DCA (used in \cite{Deeppocket} and \cite{DeepSurf}), instead of Distance from Center to Center (DCC, used in \cite{Deeppocket} and \cite{Kalasanty}). DCC, unlike DCA, computes the distance between the predicted binding site center and the center of the binding pocket associated to the ligand. Although this may be semantically more adequate than DCA, DCC has a critical limitation---it can be measured only when the binding pocket associated to the ligand is manually annotated or extracted successfully by a computational means. Indeed, in \cite{Kalasanty}, this fact limited the evaluation on a non-annotated dataset to the ligands whose associated binding pockets could be successfully extracted by an external software. 

\subsection{IOU} 
This metric is analogous to a metric used in \cite{Kalasanty} to evaluate the \textit{segmentation} ability of the model, but differs from it in two aspects. Firstly, we defined the "closest ligands" in terms of DCA, instead of DCC. Secondly, we compared the shape of the predicted binding site and the actual binding site in terms of the protein residues, instead of the segmented 3D volume as in \cite{Kalasanty}. The last choice reflects our emphasis on the task of BRI, which has its own applications that cannot be accomplished by volume segmentation, as explained in the introduction. 

\subsection{conditional IOU}
This metric is analogous to a metric used in \cite{Deeppocket}. However, their version computed the IOU out of predictions made from Fpocket-generated binding sites that are closest to the ligands, instead of binding sites that their BSD model predicted. Although this is a reasonable approach, other baseline methods that are not based on Fpocket cannot be assessed in this way.

\bibliographystyle{alpha}
\bibliography{suppl_info}
	
%------------------------------------------------------